\PassOptionsToPackage{font=small}{caption}

\documentclass[journal]{IEEEtran}
\usepackage{cite}
\usepackage{multirow}
\usepackage{subfig}
\usepackage{graphicx}

\newcounter{subcopyrightbox@save}
\usepackage{color, url}
\usepackage{xspace} 
\usepackage{mathrsfs}
\usepackage{float}
\usepackage{tabularx}

\newcommand{\argmax}{\operatornamewithlimits{argmax}}
\usepackage{amsmath}

\newcommand{\myparatight}[1]{\smallskip\noindent{\bf {#1}:}~}

\begin{document}

\title{SybilBelief: A Semi-supervised Learning Approach for Structure-based Sybil Detection}

\author{Neil Zhenqiang Gong,~\IEEEmembership{Student Member, IEEE}, Mario Frank, and Prateek Mittal,~\IEEEmembership{Member, IEEE}
\thanks{N. Z. Gong and M. Frank are with the Electrical Engineering and Computer Science Department, University of California at Berkeley, Berkeley, CA, 94720 USA, e-mail: (neilz.gong@berkeley.edu; mail2mf@gmx.de).}
\thanks{P. Mittal is with Department of Electrical Engineering, Princeton University, Princeton, NJ 08544, email: (pmittal@princeton.edu).}
\thanks{We would like to thank Dawn Song for insightful discussions. This research was supported by Intel through the ISTC for Secure Computing and by the Swiss National Science Foundation (SNSF), grant no.~138117.}
}


\maketitle

\begin{abstract}
Sybil attacks are a fundamental threat to the security of distributed systems. Recently, there has been a growing interest in leveraging social networks to mitigate Sybil attacks. However, the existing approaches suffer from one or more drawbacks, including bootstrapping
from either only known benign or known Sybil nodes, failing to tolerate noise in their prior knowledge about known benign or Sybil nodes, and not being scalable.  

In this work, we aim to overcome these drawbacks. Towards this goal, we introduce \emph{SybilBelief}, a semi-supervised learning framework, to detect Sybil nodes.  SybilBelief takes a social network of the nodes in the system,  a small set of known benign nodes, and, optionally, a small set of known Sybils as input. Then SybilBelief propagates the label information from the known benign and/or Sybil nodes to the remaining nodes in the system. 

We evaluate SybilBelief using both synthetic and real world social network 
topologies. We show that SybilBelief is able to accurately identify Sybil 
nodes with low false positive rates and low false negative rates. SybilBelief is resilient to noise in 
our prior knowledge about known benign and Sybil nodes. Moreover, SybilBelief performs orders of magnitudes better than existing Sybil classification mechanisms and significantly better than existing Sybil ranking mechanisms. 
\end{abstract}

\IEEEpeerreviewmaketitle

\section{Introduction}
\label{sec:intro}

Sybil attacks, where a single entity emulates the  behavior of multiple users,  form a fundamental threat to the security of distributed systems~\cite{sybil}. Example systems include peer-to-peer networks, email, reputation systems, and online social networks. For instance, in  
2012 it was reported that 83 million out of 900 million Facebook accounts are Sybils~\cite{Facebooksybil}. Sybil accounts in online social networks are used for criminal activities such as spreading spam or malware~\cite{Thomas11}, stealing other users' private information~\cite{Bilge09,Fong11}, and  manipulating web search results via ``+1" or ``like" clicks~\cite{search}. 

Traditionally, Sybil defenses require users to present trusted identities issued by certification authorities.  However, such approaches violate the open nature that underlies the success of these distributed systems~\cite{Viswanath10}. Recently, there has been a growing interest in leveraging  social networks to mitigate Sybil attacks~\cite{Yu06,Yu08,Danezis09,Tran09,Viswanath10,Tran11,Cao12,Yang12-spam,Alvisi13}. These schemes are based on the observation that, although an attacker can create arbitrary Sybil users and social connections among themselves, he or she can only establish a limited number of social connections to benign users.  As a result, Sybil users tend to form a community structure among themselves, which  enables a large number of Sybil users to integrate into the system.  Note that it is crucial to obtain social connections that represent trust relationships between users, otherwise the structure-based Sybil detection mechanisms have limited detection accuracy. See Section~\ref{social_model} for more discussions.  

However, existing structure-based approaches suffer from one or more of the following drawbacks: (1) they can bootstrap from either only known benign~\cite{Yu06,Yu08,Danezis09,Tran11} or known Sybil nodes~\cite{Yang12-spam}, limiting their detection accuracy (see Section~\ref{eva_syn}), (2) they cannot tolerate noise in their prior knowledge about known benign~\cite{Cao12} or Sybil nodes~\cite{Yang12-spam}, and (3) they are not scalable~\cite{Yu06,Yu08,Danezis09,Tran09,Viswanath10,Tran11}.  

To overcome these drawbacks, we recast the problem of finding Sybil users  as a semi-supervised learning problem, where the goal is to propagate reputations from a small set of known benign and/or Sybil  users to other users along the social connections between them. More specifically, we first associate a binary random variable with each user in the system; such random variable represents the label (i.e., benign or Sybil) of the user. Second, we model the social network between users in the system as a pairwise Markov Random Field, which defines a joint probability distribution for these binary random variables. Third,  given a set of known benign and/or Sybil users, we infer the posterior probability of a user being benign, which is treated as the reputation of the user.  For efficient inference of the posterior probability, we couple our framework with Loopy Belief Propagation~\cite{Pearl88}, an iterative algorithm for inference on probabilistic graphical models.

We extensively  evaluate the influence of various factors including parameter settings in the SybilBelief, the number of labels, and label noises on the performance of SybilBelief. For instance, we find that SybilBelief is relatively robust to parameter settings, SybilBelief requires one label per community, and SybilBelief can tolerate 49\% of labels to be incorrect in some cases.   In addition, we compare SybilBelief with state-of-the-art Sybil classification and ranking approaches on real-world social network topologies. Our results demonstrate that SybilBelief performs orders of magnitude better than previous Sybil classification mechanisms and significantly better than  previous Sybil ranking mechanisms.
Finally, SybilBelief proves to be more resilient to noise in our prior knowledge about known benign users and known Sybil users.

In summary, our work makes the following contributions: 
\begin{itemize}
\item We propose SybilBelief, a semi-supervised learning framework, to perform both Sybil classification and Sybil ranking. SybilBelief overcomes a number of drawbacks of previous work.

\item We extensively evaluate the impact of various factors including parameter settings in SybilBelief, the number of labels, and label noise on the performance of SybilBelief using synthetic social networks. For instance, we find that SybilBelief is relatively robust to parameter settings, SybilBelief requires one label per community, and SybilBelief can tolerate 49\% of labels to be incorrect in some cases.  

\item We  demonstrate, via evaluations on real-world social networks, that SybilBelief performs orders of magnitude better than previous Sybil classification mechanisms and significantly better than  previous Sybil ranking mechanisms. Moreover, SybilBelief is more resilient to label noise, i.e., partially corrupted prior knowledge about known benign users and known Sybil users.

\end{itemize}


\section{Problem Definition}  
We formally define the Sybil detection problem. Specifically, we first introduce the social network model. Then we introduce a few design goals.  

\begin{figure}
\centering
\includegraphics[width=0.3\textwidth, height=1.5in]{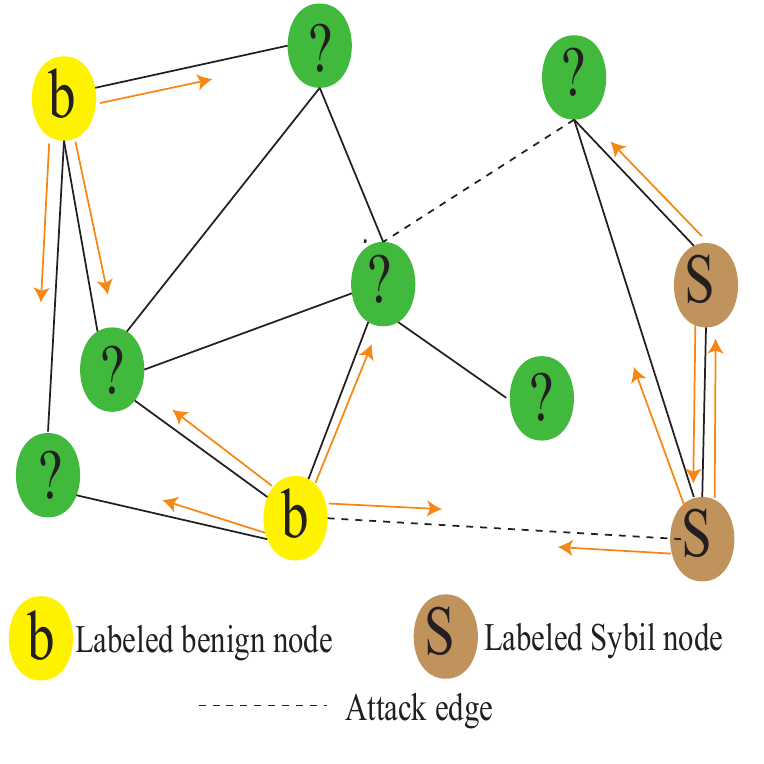}
\caption{The propagation in SybilBelief. Given a set of labeled nodes, we want to infer the labels of the remaining nodes. SybilBelief iteratively propagates the label information from the labeled nodes to their neighbors.}
\label{sybil}
\vspace{-4mm}
\end{figure}

\subsection{Social Network Model}
\label{social_model}
Let us consider an undirected social network $G=(V, E)$, where a node $v\in V$ represents a user in the system and an edge $(u, v)\in E$ indicates that the users $u\in V$ and $v\in V$ are socially connected. In an ideal setting, $G$ represents a weighted network of trust relationships between users, where the edge weights represent the levels of trust between users~\cite{Mohaisen11}. Each node is either \emph{benign} or \emph{Sybil}.

Figure~\ref{sybil} illustrates a Sybil attack. We denote the subnetwork including the benign nodes and the edges between them as the \emph{benign region}, denote the subnetwork including the Sybils and edges between them as the \emph{Sybil region}, and denote the edges between the two regions as \emph{attack edges}. Note that the benign region could consist of multiple communities and we will evaluate their impact on Sybil detections in Section~\ref{sec:exp_sb}. 

{An attacker could obtain attack edges via spoofing benign nodes to link to Sybils or  compromising benign nodes, which turns the edges between the compromised benign nodes and other benign nodes to attack edges. Compromised benign nodes are treated as Sybils, and  they could be those whose credentials are available to the attacker or \emph{front peers}~\cite{Wang10Poisonedwater} who collude with Sybils. } 


One fundamental assumption underlying the structure-based Sybil detections is that the benign region and the Sybil region are sparsely connected (i.e., the number of attack edges is small), compared to the connections among themselves. We notice that this assumption is equivalent to assuming that the social networks follow \emph{homophily}, i.e., two linked nodes tend to have the same label. For an extreme example, if the benign region and the Sybil region are isolated from each other, then the network has perfect homophily, i.e., every two linked nodes have the same label. As we will see in Section~\ref{sec:model}, the concept of homophily can better help us incorporate both known benign and Sybil nodes because it explicitly models labels of nodes.

Note that, it is crucial to obtain social networks that satisfy the homophily assumption. Otherwise the detection accuracies of  structure-based approaches  are limited. For instance,  Yang et al.~\cite{Yang11-sybil} showed that the \emph{friendship network} in RenRen, the largest online social networking site in China, does not satisfy this assumption, and thus structure-based approaches should not be applied to such friendship networks. However, Cao et al.~\cite{Cao12} found that the invitation-based friendship network in Tuenti, the leading online social network in Spain, satisfies the homophily assumption, and thus their Sybil ranking mechanism achieves reasonably good performance.  {In general,  online social network operators can obtain social networks that satisfy homophily via two methods. One method is to approximate \emph{trust} relationships between users through looking into user interactions~\cite{wilson:eurosys09}, inferring tie strengths~\cite{gilbert:chi09}, and asking users to rate their social contacts~\cite{sybildefender}. The other method is to preprocess the networks so that they are suitable for structure-based approaches. In particular, operators could first detect and remove compromised benign nodes (e.g., front peers)~\cite{Egele13,Wang10Poisonedwater}, which decreases the number of attack edges and increases the homophily. Moreover, Alvisi et al.~\cite{Alvisi13} showed that some simple detectors might enforce the social networks to be suitable for structure-based Sybil defenses if the attack edges are established randomly~\cite{Alvisi13}.}





\subsection{Design Goals}
Our goal is to detect Sybils in a system via taking a social network between the nodes in the system,  a small set of known benign nodes, and (optionally) a small set of known Sybils as input. Specifically, we have the following design goals.


{\em 1. Sybil classification/ranking:} Our goal is to 
design a mechanism that can either 
classify nodes into benign and Sybil  or that can rank all nodes in descending order of being benign.

{\em 2. Incorporating known labels:} In many settings, we already know that \emph{some} users are benign and that some users are Sybil. For instance,  in Twitter, verified users can be treated as known benign labels and users spreading spam or malware can be treated as known Sybil labels.
To improve overall accuracy of the system, the mechanism should have the ability to incorporate information about both known benign and known Sybil labels. 
It is important that the mechanism should not \emph{require} information about 
known Sybil labels, but if such information is available, then it should have 
the ability to use it. This is because in some scenarios, for example when 
none of the Sybils have performed an attack yet, we might not have 
known information about any Sybil node.

{\em 3. Tolerating label noise:} While incorporating information about 
known benign or known Sybil labels, it is important that the mechanism 
is resilient to noise in our prior knowledge about these labels. For example, 
an adversary could compromise the account of a known benign user, or could 
get a Sybil user whitelisted.  We 
target a mechanism that is resilient 
when a minority fraction of known 
labels are incorrect.

{\em 4. Scalability:} Many distributed systems (e.g., online social networks,  reputation systems) have hundreds of millions of 
users and billions of edges. Thus, for real world applicability, 
the computational complexity of the mechanism should be low, and the 
mechanism should also be parallelizable. 

Requirements 2, 3, and 4 distinguish our framework from prior work. Sybil classification approaches such 
as SybilLimit~\cite{Yu08} and SybilInfer~\cite{Danezis09} do not incorporate 
information about known Sybil labels (limiting detection 
accuracy, as shown in Section~\ref{eva_syn}), are not resilient to label noise\footnote{These Sybil classification mechanisms only incorporate one labeled benign node, which makes them not resilient to label noise.}, and are not scalable. 
Sybil ranking approaches such 
as SybilRank~\cite{Cao12} and CIA~\cite{Yang12-spam} incorporate information about 
either known benign or known Sybil labels, but not both. They are also not resilient 
to label noise. 


%

\section{SybilBelief model} 
\label{sec:model}
We introduce our approach SybilBelief,
which is scalable, tolerant to label noise, and able to incorporate  both known benign labels and Sybil labels.
    
\subsection{Model Overview}
To quantify the homophily in social networks, we first propose a new probabilistic local rule which
determines the reputation score for a node $v$ via aggregating its neighbors' label information. Then, we
demonstrate that this local rule can be captured by modeling social networks as
Markov Random Fields (MRFs). Specifically, each node in the network is
associated with a binary random variable whose state  could either be
\emph{benign} or  \emph{Sybil}, and MRFs define a joint probability
distribution over all such random variables. Given a set of known benign labels
and/or known Sybil labels, the posterior probabilities that nodes are  benign are used to classify or rank them.
We adopt Loopy Belief Propagation~\cite{Pearl88,Murphy99} to approximate the
posterior probabilities. Figure~\ref{sybil} illustrates how SybilBelief
iteratively propagates the beliefs/reputations from the labeled nodes to
the remaining ones.

\subsection{Our Probabilistic Local Rule}
Recall that we have a social network $G=(V,E)$ of the nodes in the system. Each node  can have two states, i.e., benign or Sybil. Thus, we associate a binary random variable  $x_v\in\{-1, 1\}$ with each node.  $x_v=+1$ means that node $v$ is a benign node and $x_v=-1$ indicates that node $v$ is 
Sybil.  
 In the following, we use $x_A$ to represent the set of random variables associated with the nodes in the set  $A$. Moreover, we use $\bar{x}_A$ to denote the observed values of these random variables. 
 
There might exist some prior information about a node $v$ independently from 
 all other nodes in the system. Such prior information could be the content generated by $v$ or its behavior. We model the prior belief of $v$ being benign as follows:   
\begin{align}
P(x_v=+1)= \frac{1}{1+\text{exp}(-h_v)} \ ,
\end{align}
where $h_v$ quantifies the prior information about $v$. More specifically, $h_v>0$ encodes the scenario in which $v$ is more likely to be benign; $h_v<0$ encodes the opposite  scenario; $h_v=0$ means prior information is not helpful to determine $v$'s state.  

We now introduce $\Gamma_v=\{u|(u,v)\in E\}$, the set of  $v$'s neighbors in the social network, and their respective states $\bar{x}_{\Gamma_v}$.
When these states are known, the probability of $v$ to be benign is modeled as
\begin{align}
\label{local_rule}
P(x_v=+1|\bar{x}_{\Gamma_v})=\frac{1}{1+\text{exp}(-\sum_{u\in \Gamma_v}J_{uv}\bar{x}_u - h_v)} \ ,
\end{align}
where $J_{uv}$ is the coupling strength between $u$ and $v$. Specifically, $J_{uv}>0$ means $u$ and $v$ tend to have the same state;  $J_{uv}<0$ indicates $u$ and $v$ tend to have opposite states; and $J_{uv}=0$ indicates that there is no coupling between them. In practice, these coupling strengths can encode trust levels between nodes. 

Note that our local rule in Equation~\ref{local_rule} incorporates the homophily assumption via setting $J_{uv}>0$.

\subsection{A Pairwise Markov Random Field}
We find that the probabilistic local rule introduced in the previous section can be applied by modeling the social network as a pairwise Markov Random Field (MRF). A MRF defines a joint probability distribution for binary random variables associated with all the nodes in the network. Specifically, a MRF is specified with a \emph{node potential} for each node $v$, which incorporates prior knowledge about $v$, and with an \emph{edge potential} for each edge $(u,v)$, which represents correlations between $u$ and $v$.    In the context of Sybil detection, we define a node potential $\phi_v(x_v)$ for the node $v$ as
\begin{displaymath}
\phi_v(x_v):=
\begin{cases} 
\theta_v  & \text{if } x_v = 1  \\ 
1-\theta_v  &  \text{if } x_v = -1
\end{cases} 
\end{displaymath}
and an \emph{edge potential  $\varphi_{uv}(x_u, x_v)$} for the edge $(u,v)$ as
\begin{displaymath}
\varphi_{uv}(x_u, x_v):=
\begin{cases} 
w_{uv} & \text{if } x_u x_v = 1  \\  
1 - w_{uv} &  \text{if } x_u x_v = -1 \ ,
\end{cases}
\end{displaymath}
where $\theta_v\!:=\!(1\!+\!\mathrm{exp}\{-h_v\})^{-1}$ and $w_{uv}\!:=\!(\!1+\!\mathrm{exp}\{-J_{uv}\})^{-1}$. 
Then, the following MRF  satisfies the probabilistic local rule.
$$P(x_V)= \frac{1}{Z}\prod_{v\in V} \phi_v(x_v) \prod_{(u, v)\in E} \varphi_{uv}(x_u, x_v)\ ,$$
where $Z=\sum_{x_V}\prod_{v\in V} \phi_v(x_v) \prod_{(u, v)\in E} \varphi_{uv}(x_u, x_v)$ is called the partition function and normalizes the probabilities.

The node potential $\phi_v(x_v)$ incorporates our prior knowledge about node $v$.  Specifically, setting $\theta_v > 0.5$ assigns a higher probability to node $v$ of being benign than Sybil. Setting $\theta_v < 0.5$ models the opposite and $\theta_v = 0.5$  is indifferent between the two states. 
This is the mechanism through which our framework can incorporate given benign or Sybil labels.

The  edge potential $\varphi_{uv}(x_u, x_v)$ encodes the coupling strength of two linked nodes $u$ and $v$. The larger $w_{uv}$ is, the stronger the model favors $u$ and $v$ to have the same state.   More precisely,  $w_{uv} > 0.5$ means connected nodes tend to have the same state; $w_{uv} < 0.5$ means connected nodes tend to have opposite states; $w_{uv}=0.5$ encodes the scenario where there is no coupling between $u$ and $v$.  Note that setting $w_{uv} > 0.5$ encodes our homophily assumption.

\subsection{Detecting Sybils}
First, we modify the above MRF to incorporate known labels. Then we use the modified model to perform both Sybil classification and Sybil ranking.

\myparatight{Leveraging given labels} Suppose we observe states for a set of nodes, and let us denote them as $L\subseteq V$. We use these labels to set the corresponding parameters in the MRF as follows.  
For any unlabeled node $v$, i.e., $v\notin L$, we set $\theta_v=0.5$. 
Moreover, our inference model (i.e., Equation~4) does not rely on the prior beliefs of those labeled nodes. So we can set $\theta_v$ to be any positive value, where  $v\in L$.
The coupling strength parameter is set as $w_{uv}=w>0.5$ for any edge $(u,v)$ to model the homophily assumption\footnote{In principle, the coupling strength parameters can incorporate the trust levels between nodes, and thus they can be different for different edges.}.  



To be convenient in our later analysis, we define \emph{evidence potentials} using  known labels as follows:
 \begin{align}
\phi_v^L(x_v)=
\begin{cases} 
\phi_v(x_v) \delta_{x_v \bar{x}_v}  & \text{if } v \in L  \\ 
\phi_v(x_v)  &  \text{if } v \notin L \ ,
\end{cases}
\end{align}
where $\bar{x}_v$ is the known label of node $v$, and $\delta_{x_v \bar{x}_v}$ is the Kronecker delta function, i.e., $\delta_{x_v \bar{x}_v}=1$ if $x_v= \bar{x}_v$, otherwise $\delta_{x_v \bar{x}_v}=0$.

In our modified model, given the labeled node set $L$, we can compute the posterior distribution for each unlabeled node $v$. Specifically, we have
\begin{align}
P(x_v|\bar{x}_L) &= \sum_{x_{V/v}} P(x_V|\bar{x}_L) \label{eq_condProbs}
\\
&= \frac{1}{Z^L}  \sum_{x_{V/v}} \prod_{v\in V} \phi_v^L(x_v) \prod_{(u, v)\in E} \varphi_{uv}(x_u, x_v)
\nonumber
\end{align}

The posterior probabilities $P(x_v=+1|\bar{x}_L)$ that $v$ is benign given observed nodes $L$ are used to \emph{classify} or \emph{rank} them. 

\myparatight{Sybil classification} We map the Sybil classification problem as the following inference problem.
$$y_v =\argmax_{i\in\{-1,1\}} P(x_v=i|\bar{x}_L) $$
where $y_v$ is the inferred label of $v$, i.e.,  $y_v=1$ indicates that $v$ is benign, otherwise $v$ is Sybil. 

\myparatight{Sybil ranking} We use the posterior probabilities $P(x_v=+1|\bar{x}_L)$ to rank all the unlabeled nodes. 
 

\myparatight{Boosting} We use a boosting strategy for our algorithm if either only labeled benign nodes or only labeled Sybil nodes are given. Since both cases 
 are algorithmically equivalent, we take the first case as an example to illustrate our strategy.  

We first sample some nodes uniformly at random from the entire system, and we treat them as labeled Sybil nodes. Then we compute the posterior distribution of every unlabeled node. In each such process, we get a posterior distribution for every node. Furthermore, we repeat this process $K$ times. Thus, we get $K$ posterior distributions for every node, which are denoted as $P_i(x_v|\bar{x}_{L_i})$, where $i=1,2,\cdots,K$.  We aggregate the $K$ posterior distributions for every node as follows:
 \begin{align}
P(x_v=-1|\bar{x}_L)=\max_{i}P_i(x_v=-1|\bar{x}_{L_i}) \nonumber
\\
P(x_v=+1|\bar{x}_L)=1-P(x_v=-1|\bar{x}_L) \nonumber
\end{align}

The aggregated posterior distributions are then used to classify or rank nodes. This boosting strategy works because our model can update prior beliefs and therefore is robust to label noise (see Section~\ref{eva_syn} and~\ref{sec:exp_sb}). In each boosting trial, if some of the sampled nodes are  
true Sybil nodes, then this trial can detect a subset of Sybil nodes.  Due to the robustness to label noise, even if some sampled nodes are actually benign, the propagation of Sybil beliefs among the benign region is limited once 
the number of such sampled nodes is smaller than the number of labeled benign nodes. Thus in our experiments, we limit the number of sampled Sybil nodes in the boosting process 
by the number of labeled benign nodes.     

\section{SybilBelief Learning Algorithm}
\label{sec_algo}
Our Sybil classification and ranking mechanisms rely on the computation of the posterior distributions given in Equation~\ref{eq_condProbs}. Generally, there are two major ways to infer such posterior distributions: \emph{sampling} and \emph{variational inference}. We adopt variational inference to learn the posterior distributions since it is more scalable than sampling approaches such as Gibbs sampling. Specifically, we adopt Loopy Belief Propagation (LBP) to calculate the posterior distributions for each node.

\myparatight{Loopy Belief Propagation (LBP)~\cite{Pearl88}} 
The basic step in LBP is to pass messages between neighboring nodes in the system.  Message $m_{uv}^{(t)}(x_v)$ sent from $u$ to $v$ in the $t$th iteration is 
$$m_{uv}^{(t)}(x_v) = \sum_{x_u} \phi_u^L(x_u) \varphi_{uv}(x_u, x_v) \prod_{k \in \Gamma(u)/v} m_{ku}^{(t-1)}(x_u)$$ 
Here, $\Gamma(u)/v$ is the set of all neighbors of $u$, except the receiver node $v$. This encodes that each node 
 forwards a product over  incoming messages of the last iteration and adapts this message to the respective receiver based on the coupling strength with the receiver. 

For social networks without loops (i.e., for trees), LBP is guaranteed to converge and to compute the exact posterior distribution. For networks with loops, LBP approximates the posterior probability distribution without convergence guarantees. However, in practical applications and benchmarks in the machine learning literature~\cite{Murphy99}, LBP has demonstrated good results and is, today, a widely used technique.

\myparatight{Stopping condition} 
The message passing iterations stop when the changes of messages  become negligible (e.g.,  L1 distance of changes becomes smaller than $10^{-3}$) or the number of iterations exceeds a predefined threshold.  
After stopping, we estimate the posterior probability distribution $P(x_v|\bar{x}_L)$  by
$$P(x_v|\bar{x}_L)\propto \phi_v^L(x_v) \prod_{k \in \Gamma(v)} m_{kv}^{(t)}(x_v)$$

%
%

\myparatight{Scalability} The complexity of one LBP iteration is $O(m)$, where $m$ is the number of edges. So the total complexity is $O(m*d)$, where $d$ is the number of LBP iterations. Note that social networks are often sparse graphs~\cite{Mislove07,Gong12-imc}.  Thus we have $O(m*d)=O(n*d)$, where $n$ is the number of nodes.  Moreover, we find that setting $d$ to be 10 already achieves good results  in our experiments.  Furthermore, LBP can be easily parallelized. Specifically, we can distribute nodes in the system to multiple  processors or computer nodes, each of which collects messages for nodes assigned to them.

\section{Evaluating SybilBelief}
\label{sec:exp_sb}
We evaluate the influence of various factors including parameter settings in SybilBelief, 
the number of labels, label sites, label noises, mixing time of the social networks, and scenarios where only labeled benign or  Sybil nodes are observed, on the performance of SybilBelief. Since these experiments require social networks with various sizes, we will use well-known network generators (e.g., Erdos-Renyi model (ER)~\cite{Erdos59} 
and the Preferential Attachment (PA) model~\cite{Barabasi99}) to synthesize both the benign region and the Sybil region.
We will also study the impacts of different network generators. Furthermore, throughout these experiments, 
we view SybilBelief as a classification mechanism. 

In the basic experimental setup, we adopt the PA model to generate both the benign region and the Sybil region with 
an average degree of 10, add 500 attack edges between them uniformly at random, and 
fix the benign region to have 1000 nodes; we set the SybilBelief model 
parameters as $\theta_v=0.5$ for any node $v$, and $w_{uv}=w=0.9$ 
for any edge $(u,v)$; we assume one labeled benign node and one labeled Sybil node; the labeled benign (Sybil) 
nodes are uniformly sampled from the benign (Sybil) network. When we study the impact of one factor, we fix other factors to be the same as in the basic setup  
and vary the studied one. Moreover, all of our results reported in the following are averaged over 100 trials.

\begin{figure}[t]
\vspace{-4mm}
\centering
\subfloat[ER-ER]{\includegraphics[width=0.25\textwidth, height=1.5in]{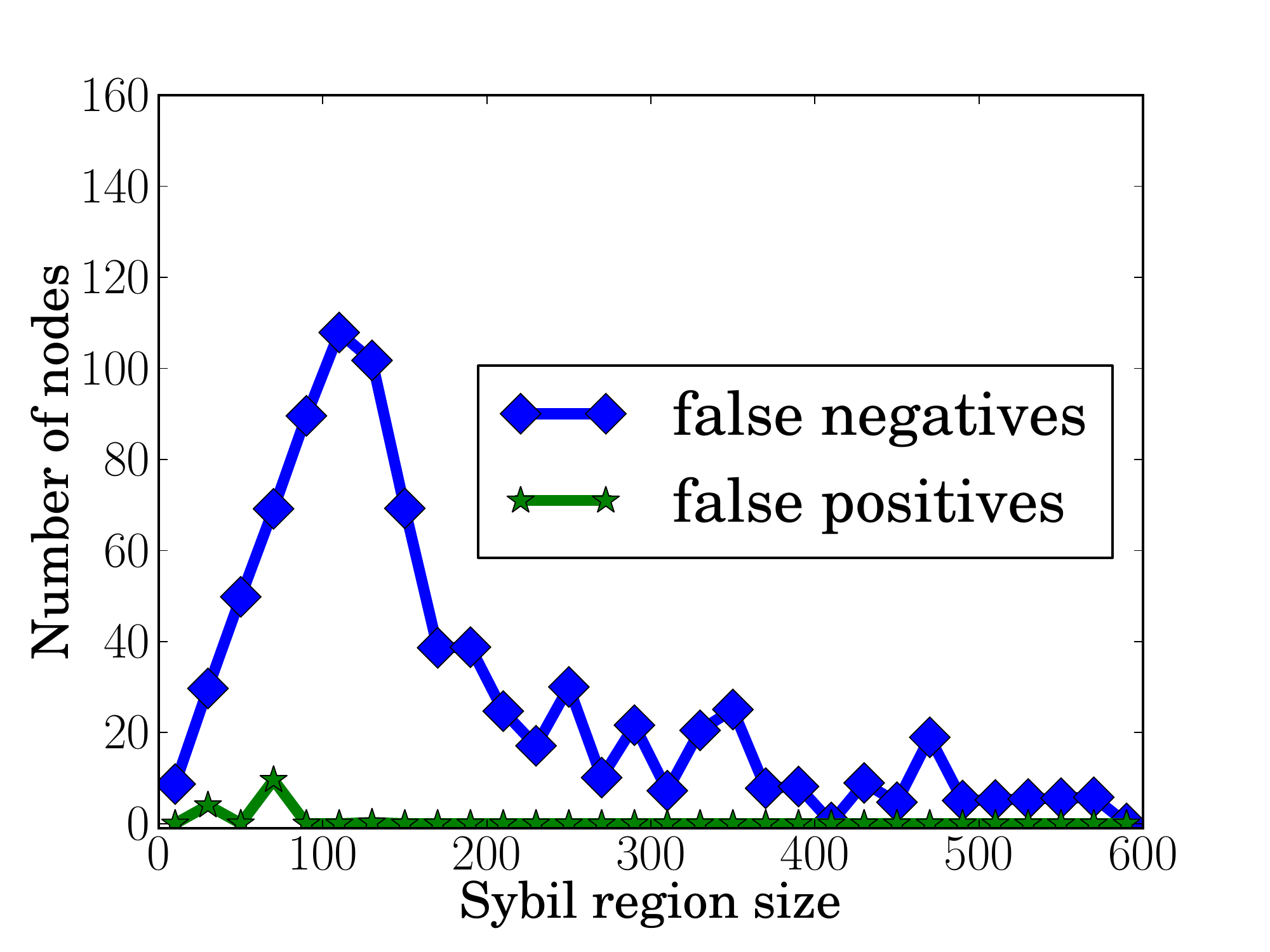}\label{accepted_sybil_random}}
\subfloat[PA-PA]{\includegraphics[width=0.25\textwidth, height=1.5in]{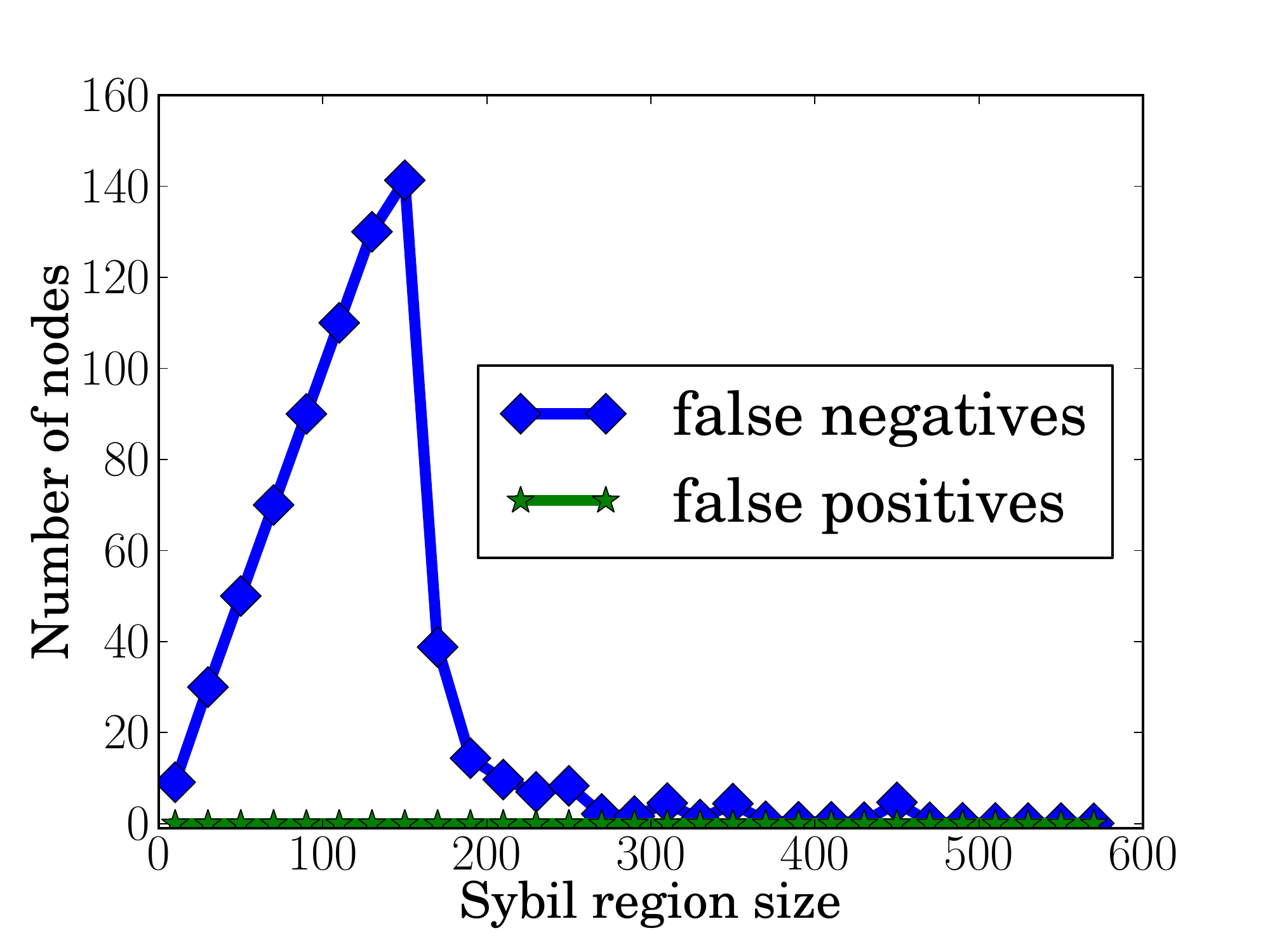}\label{accepted_sybil_pa}}
\caption{The false negatives and false positives as a function of the Sybil region size. (a) Networks are synthesized by the ER model. (b) Networks are synthesized by the PA model. There exists an optimal strategy for the attackers.}
\label{accepted_sybil}
\vspace{-4mm}
\end{figure}

\begin{figure}[t]
\centering
\subfloat[Network generator impact]{\includegraphics[width=0.25\textwidth, height=1.5in]{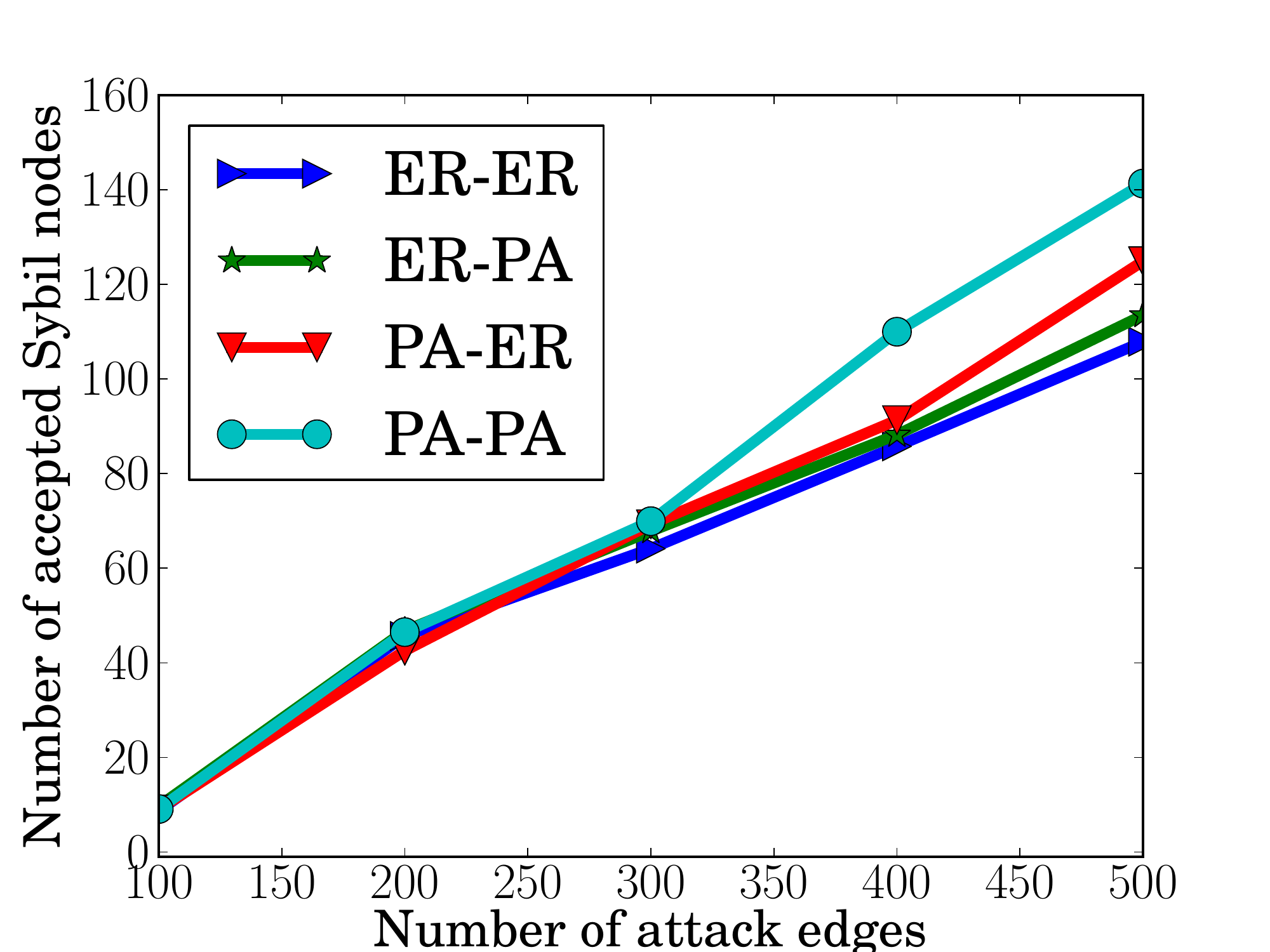}\label{explore_net_model_sybil}}
\subfloat[Label sites impact]{\includegraphics[width=0.25\textwidth, height=1.5in]{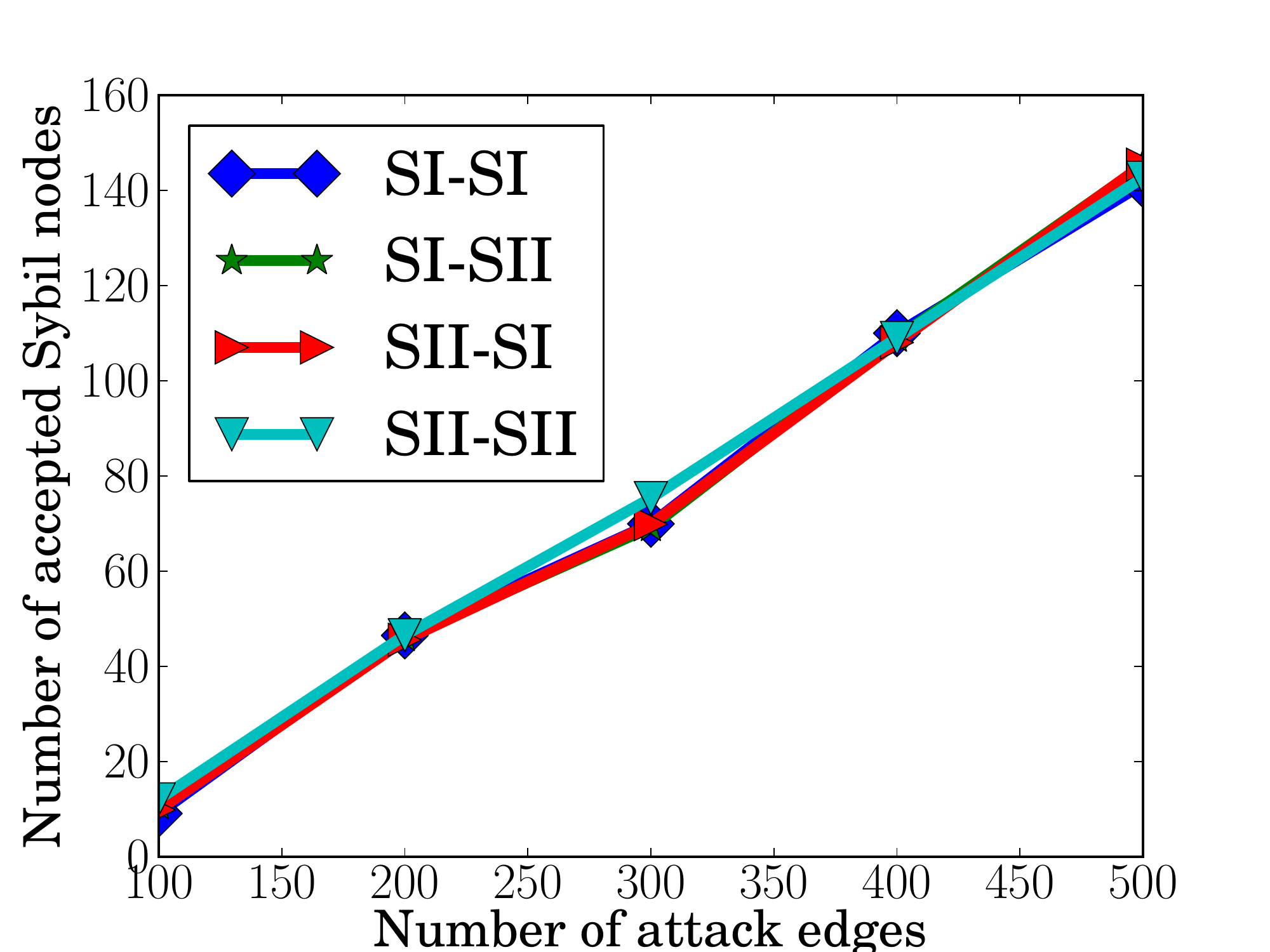}\label{explore_label_sites_sybil}}
\caption{The accepted Sybil nodes as a function of the number of attack edges. (a) Impact of different network generators. The notation $M1-M2$ means we use $M1$ to produce the benign region and $M2$ to synthesize the Sybil region. We observe that the attackers should design their Sybil regions to approach scale-free networks in order to inject more Sybils. (b) Impact of different combinations of label sites. Our algorithm SybilBelief is robust to label sites.}
\label{explore_net_model}
\vspace{-4mm}
\end{figure}

\myparatight{Impact of network generators}
We first study the impacts of network generators on the performances of SybilBelief.  To this end,  
we choose the ER network generator~\cite{Erdos59} and the PA network generator~\cite{Barabasi99}. 

For each triple (benign region, Sybil region, attack edges) setting, we can compute the 
false negatives and false positives. Figure~\ref{accepted_sybil} shows false negatives and positives 
as a function of the Sybil region  size. We find that both false negatives and false positives first 
increase and then decrease as we increase the Sybil region size. The reason for this sudden decrease phenomenon is that when the Sybil region size is bigger than some threshold, the homophily is strong enough so that SybilBelief can easily distinguish between the benign and Sybil regions.  For a (benign region, attack edges) setting, we can search for the maximum false negatives and false positives via increasing the 
Sybil region size. We denote these maximum false negatives as \emph{accepted Sybil nodes} and 
maximum false positives as \emph{rejected benign nodes}, which will be used as the metrics to 
evaluate Sybil classification systems. 

Figure~\ref{explore_net_model_sybil} shows the accepted Sybil nodes as a function of the number of 
attack edges while fixing the benign region to be the  setting in the basic setup for different network generator combinations. We find that PA-generated networks have more accepted Sybil nodes than ER-generated networks. 
Our findings imply that attackers should generate scale-free networks in order to inject more Sybils 
to the benign region. The rejected benign nodes are always less than 5 (i.e., the false positive rates 
are smaller than 0.5\% since we have 1000 benign nodes) in these experiments. We don't show them due to 
the limited space.

\myparatight{Impact of label sites}
In practice, some known labeled nodes might be 
end points of attack edges while others might be far away from attack edges.  
So one natural question is which ones to be selected as the input labels for SybilBelief. Specifically, we consider the following scenarios.

\begin{itemize}
\item {\bf SI:} Labeled benign (Sybil) nodes are not end points of attack edges. 
\item {\bf SII:} Labeled benign (Sybil) nodes are end points of attack edges. This could correspond to the scenario where sophisticated attackers obtain knowledge about the labels used in SybilBelief and establish targeted attacks.
\end{itemize}

Figure~\ref{explore_label_sites_sybil} shows the accepted Sybil nodes as a function of the number 
of attack edges for the four combinations of the label sites. We find that the label sites have no influence on the number of accepted Sybil nodes. Again, the rejected benign 
nodes are always less than 5 (i.e., the false positive rates are smaller than 0.5\%), which are 
not shown due to the limited space. Our results imply that the administrator could simply select 
benign/Sybil nodes uniformly at random as input labels for SybilBelief.

\begin{figure}[t]
\vspace{-4mm}
\centering
\subfloat[$\theta_{l}=0.50$]{\includegraphics[width=0.25\textwidth, height=1.5in]{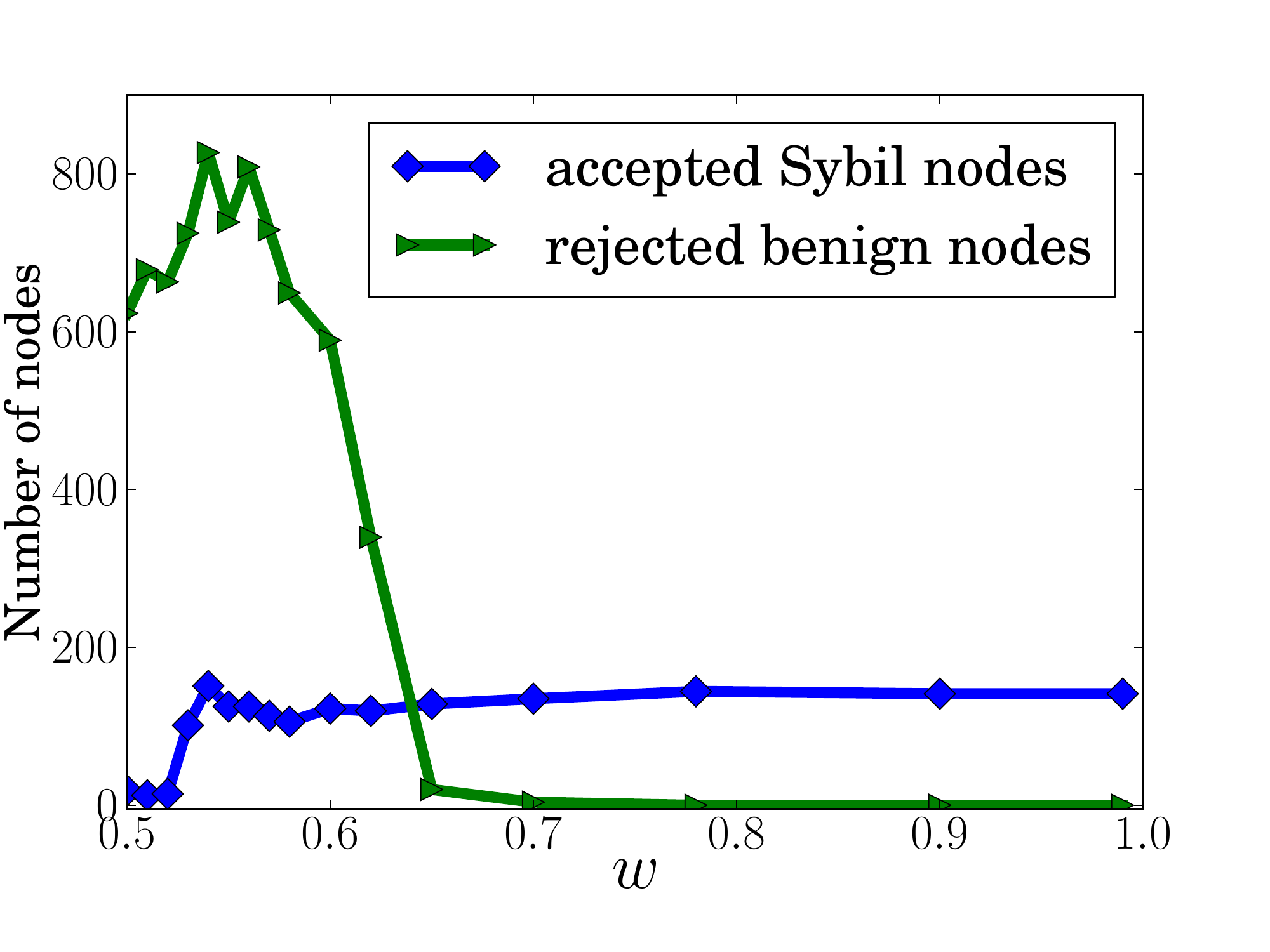}\label{explore_model_par_w}}
\subfloat[$w=0.90$]{\includegraphics[width=0.25\textwidth, height=1.5in]{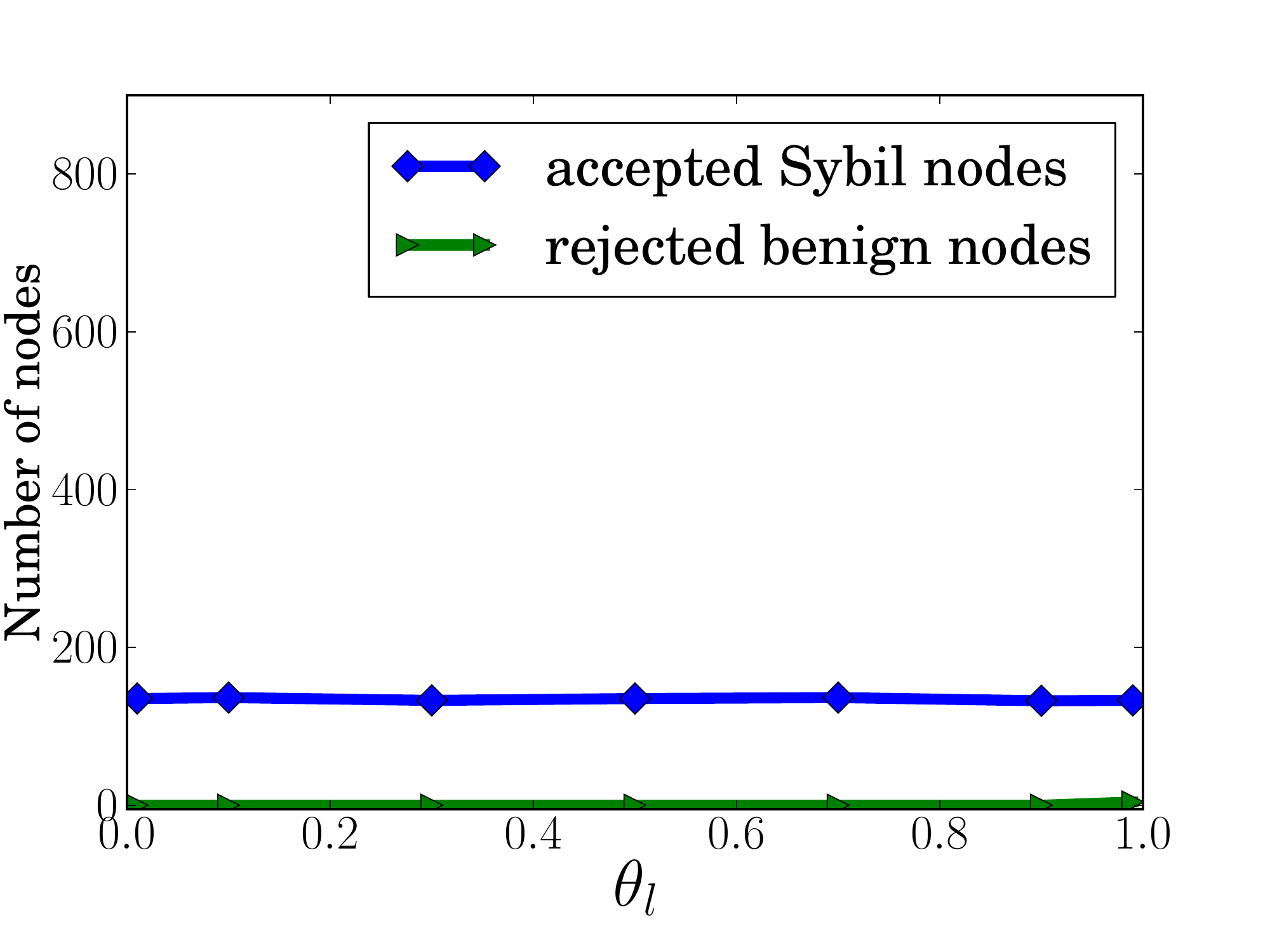}\label{explore_model_par_theta}}
\caption{The number of accepted Sybil nodes and rejected benign nodes as a function of the model parameters $w$ and $\theta_{l}$. (a) $\theta_l=0.50$ and we vary $w$. (b) $w=0.90$ and we vary $\theta_l$. We observe that there exists a phase transition point $w_0$ (e.g., $w_0\approx 0.65$ in our experiments) for the parameter $w$. SybilBelief is robust for $w>w_0$. Moreover, we confirm that SybilBelief performance is independent with $\theta_{l}$ once it's bigger than 0.}
\label{explore_model_par}
\vspace{-4mm}
\end{figure}

\myparatight{Impact of the model parameters}
In SybilBelief, there exists 
one parameter $w_{uv}$ corresponding to the homophily strength of the edge $(u,v)$.  We set $w_{uv}=w$ for all edges, and Figure~\ref{explore_model_par} illustrates the impact of $w$ on the performance of SybilBelief. To demonstrate that our model is independent with $\theta_v$ for $v \in L$,  Figure~\ref{explore_model_par} also shows the case in which we vary $\theta_l$, where $\theta_l=\theta_v$ for $v\in L$. We observe that there exists a phase transition point $w_0$ (e.g., $w_0\approx 0.65$ in these experiments) for $w$. When $w>w_0$, SybilBelief achieves a good tradeoff between accepted Sybil nodes and rejected benign nodes.

\begin{figure}[t]
\vspace{-4mm}
\centering
\subfloat[]{\includegraphics[width=0.25\textwidth, height=1.5in]{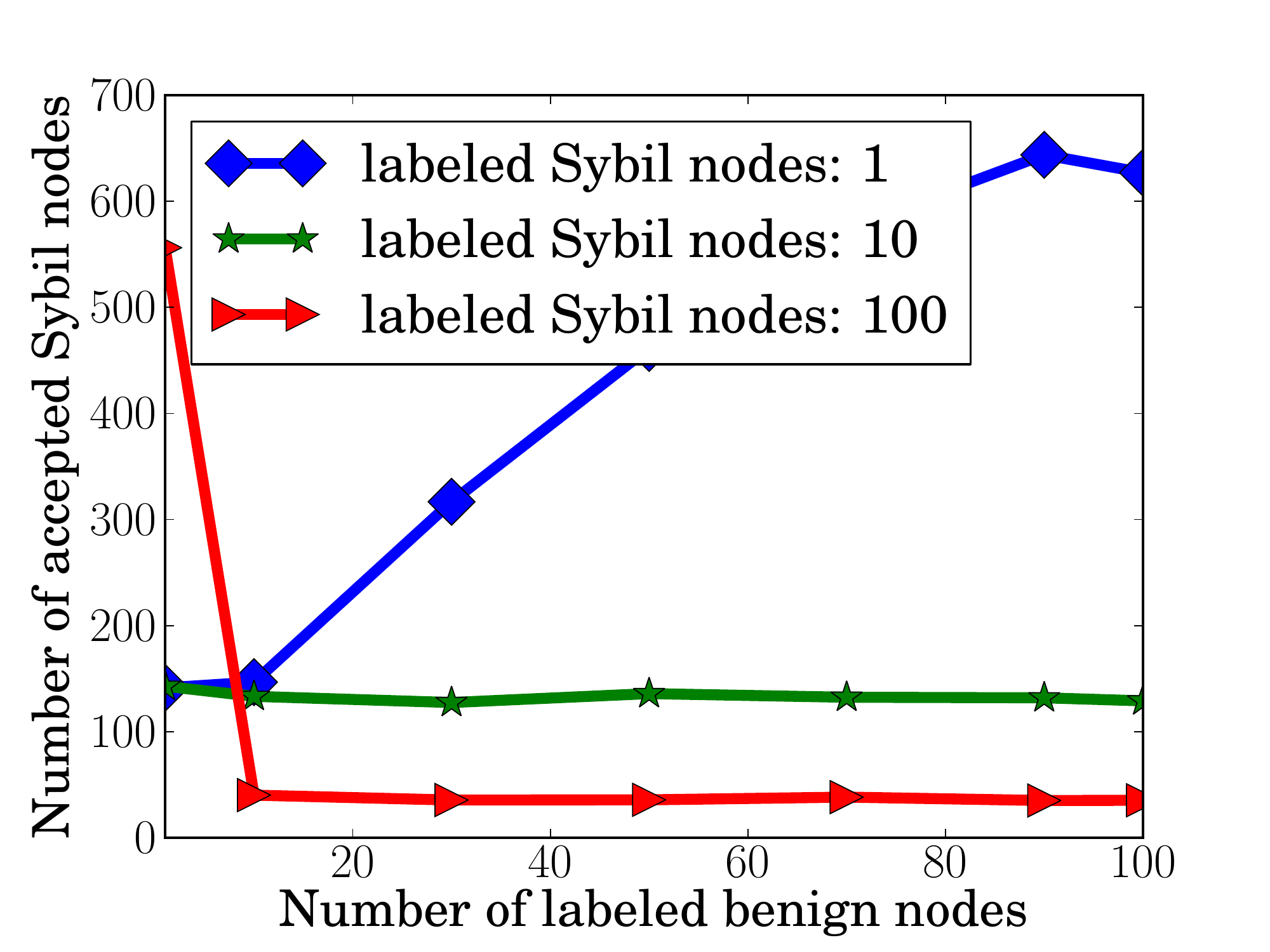}\label{explore_label_nodes_sybil_benign}}
\subfloat[]{\includegraphics[width=0.25\textwidth, height=1.5in]{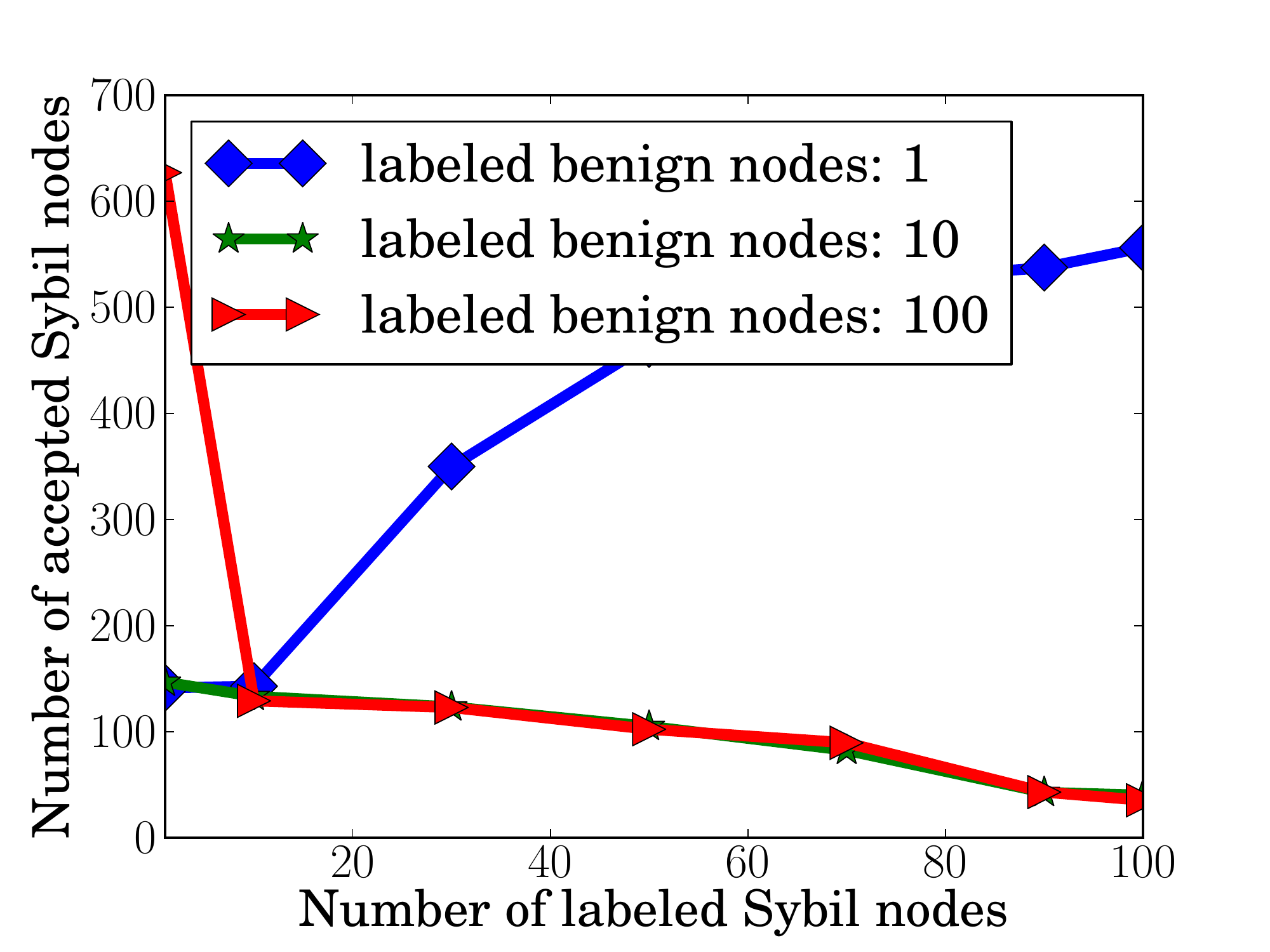}\label{explore_label_nodes_sybil_sybil}}
\caption{The number of accepted Sybil nodes as a function of the number of labeled benign and Sybil nodes. (a) The number of Sybil nodes is fixed while varying the number of benign labels. (b) The number of labeled benign nodes is fixed while increasing the number of Sybil labels. We observe that SybilBelief only requires one label per community.}
\label{explore_label_nodes}
\vspace{-4mm}
\end{figure}

\myparatight{Impact of the number of labeled nodes}
Figure~\ref{explore_label_nodes} shows the influence of various number of labels on the performance of SybilBelief. We observe that the number of accepted Sybil nodes increases dramatically when the labeled 
benign and Sybil nodes are highly imbalanced, i.e., their ratio is bigger than 10 or smaller than 0.1. Again, 
the rejected benign nodes are always less than 5 (i.e., the false positive rates are smaller than 0.5\%), which 
we don't show here due to the limited space. When the ratio is between 0.1 and 10, the number of accepted Sybil 
nodes is stable across various number of labeled benign nodes. Having more labeled Sybil nodes helps SybilBelief accept 
fewer Sybil nodes. However, these achieved margins are just the more labeled Sybils\footnote{We will not accept 
these labeled Sybil nodes.}. So we conclude that SybilBelief only needs one label for each community.

\begin{figure}[t]
\centering
\subfloat[Label noise impact]{\includegraphics[width=0.25\textwidth, height=1.5in]{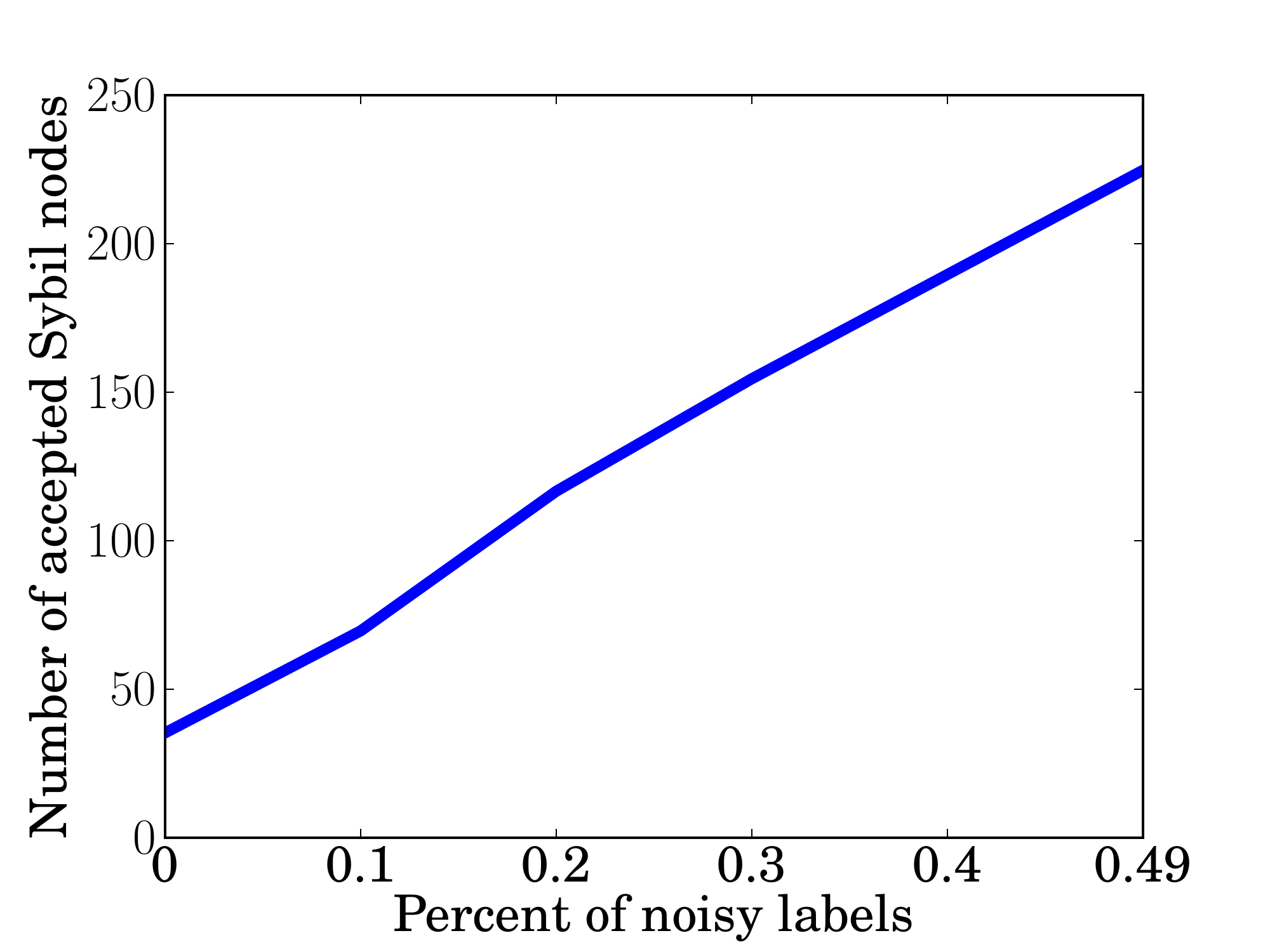}\label{explore_label_noise}}
\subfloat[Mixing time impact]{\includegraphics[width=0.25\textwidth, height=1.5in]{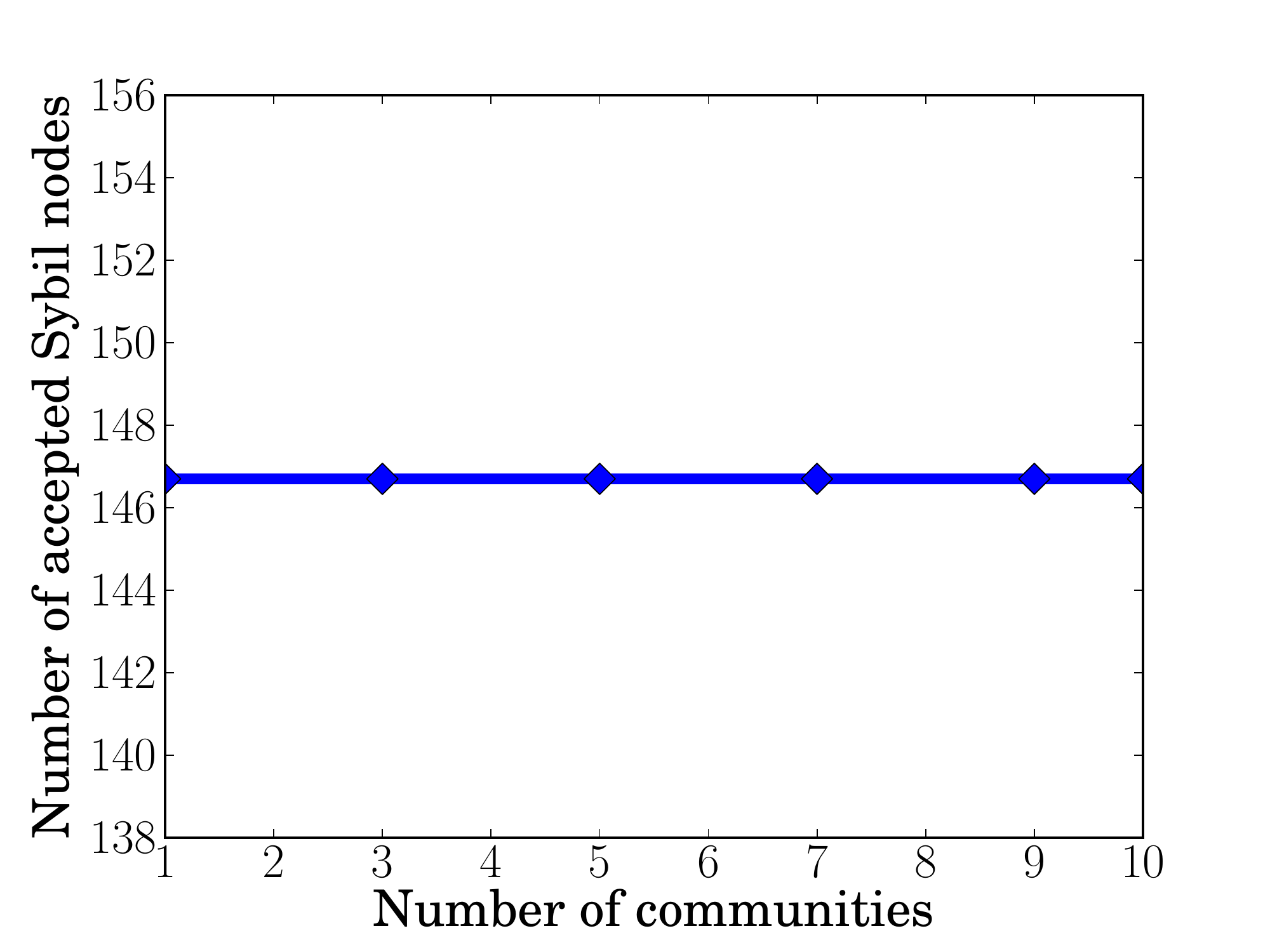}\label{mixingtime}}
\caption{Impact of label noise and community structure of the benign region on the number of accepted Sybil nodes. (a) We have 100 labeled benign and Sybil nodes. The x-axis is the percent of both noisy benign and Sybil labels. We find that SybilBelief can tolerate 49\% of labels to be incorrect. (b) The benign region consists of multiple communities. We observe that SybilBelief is robust to community structures.}
\vspace{-4mm}
\end{figure}

\myparatight{Impact of label noise}
In practice, we could use a machine learning classifier or crowdsourcing system to obtain labels.  For instance, Thomas et al.~\cite{Thomas11} used a classifier to assign labels for Twitter accounts. 
Wang et al.~\cite{Wang13} proposed to  label nodes via a crowdsourcing platform such as Amazon Mechanical 
Turk\footnote{\url{https://www.mturk.com/mturk/welcome}}. Unfortunately, labels obtained from these approaches 
often have noise. For example, labels got from crowdsourcing could have noise up to 35\%~\cite{Wang13}. Furthermore, an adversary could compromise a known benign node, or could get a Sybil node whitelisted.
Thus, one natural question is how label noise affects SybilBelief's performance.

Figure~\ref{explore_label_noise} shows the influences of such label noise. Unsurprisingly, we find that
 SybilBelief accepts more Sybil nodes and rejects more benign nodes when a larger fraction of labels are incorrect. However, with even 49\% noise\footnote{$\ge$ 50\% noise makes any algorithm accept infinite number of Sybils.}, SybilBelief 
still performs very well, i.e., SybilBelief with 49\% noise only accepts three times more Sybil nodes  than SybilBelief without noise, and  its rejected benign nodes are always less than 3 (i.e., the false positive rates are smaller than 0.3\%). 
Furthermore, we find that our algorithm can detect the incorrect labels with 100\% accuracy.

\myparatight{Impact of community structure}
Most existing Sybil detection mechanisms~\cite{Yu06,Yu08,Danezis09,Tran09,Viswanath10,Tran11,Cao12,Yang12-spam} 
rely on the assumption that the benign region is fast mixing. However, Mohaisen~\cite{mohaisen:imc10} showed that 
many real-world social networks may not be as fast-mixing as was previously thought. Furthermore, Viswanath 
et al.~\cite{Viswanath10} showed that the performance of existing Sybil detection methods decrease 
dramatically when the benign region consists of more and more communities (i.e., the mixing time is larger and larger). 

Similar to the experiment done by  Viswanath et al., we study the impact of the community structure (i.e., the mixing time) 
on the performance of SybilBelief. Specifically, for the benign region, we use PA to generate $k$ independent communities. 
For the $i$th community, we link it to the previous $i-1$ communities via 10 random edges. The Sybil region is one community 
generated by PA.  We randomly sample 1  Sybil label from the Sybil region and 10 benign labels from the benign region such that 
each community has at least 1 labeled node. Figure~\ref{mixingtime} shows the number of accepted Sybil nodes as a function of 
the number of communities in the benign region. The number of rejected benign nodes is always close to 0 and not shown here due 
to limited space. We conclude that SybilBelief is robust to community structure in the benign region.

\begin{figure}[t]
\vspace{-4mm}
\centering
\subfloat[Accepted Sybil nodes]{\includegraphics[width=0.25\textwidth, height=1.5in]{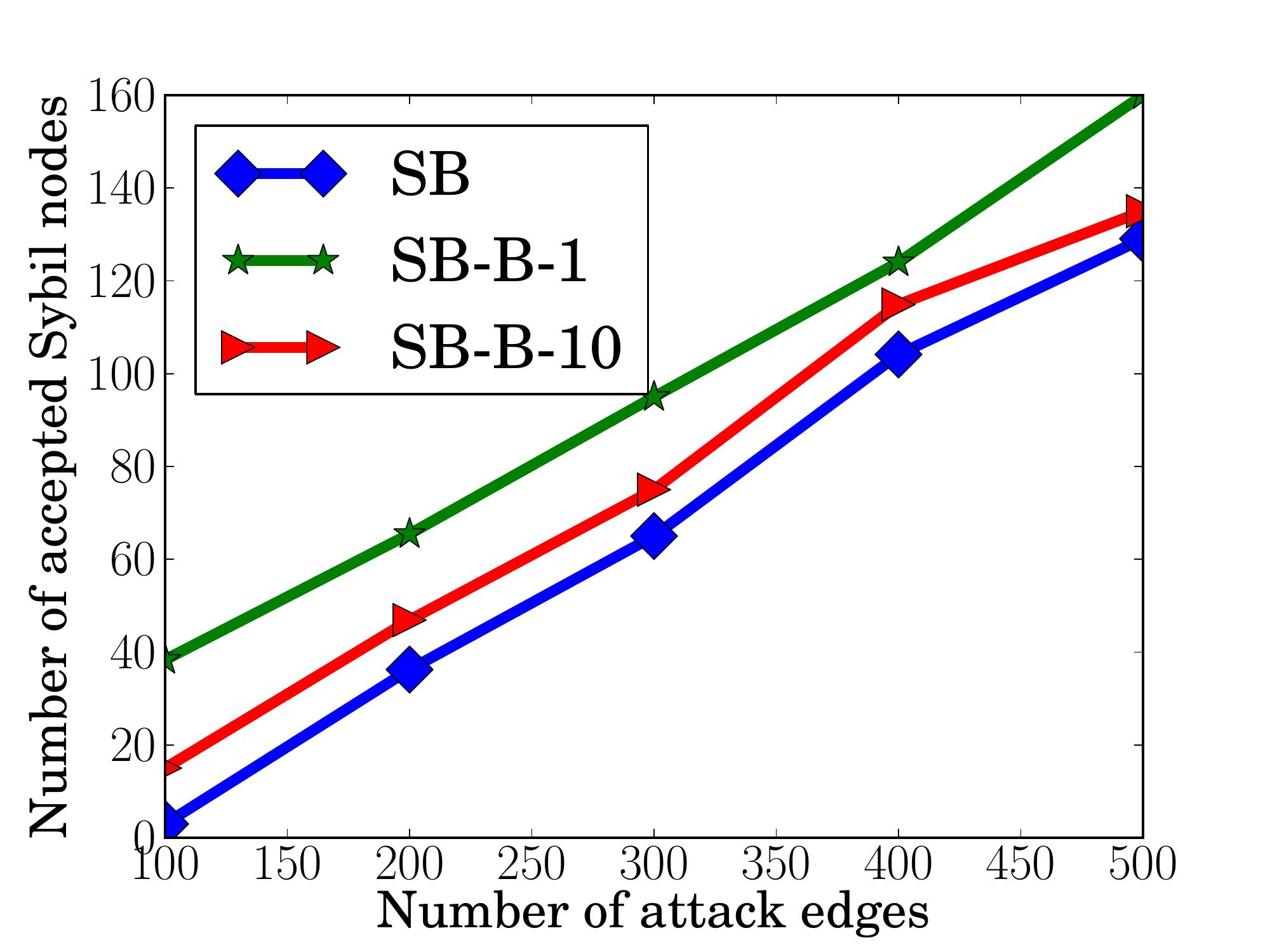}\label{explore_label_sampling_sybil}}
\subfloat[Rejected benign nodes]{\includegraphics[width=0.25\textwidth, height=1.5in]{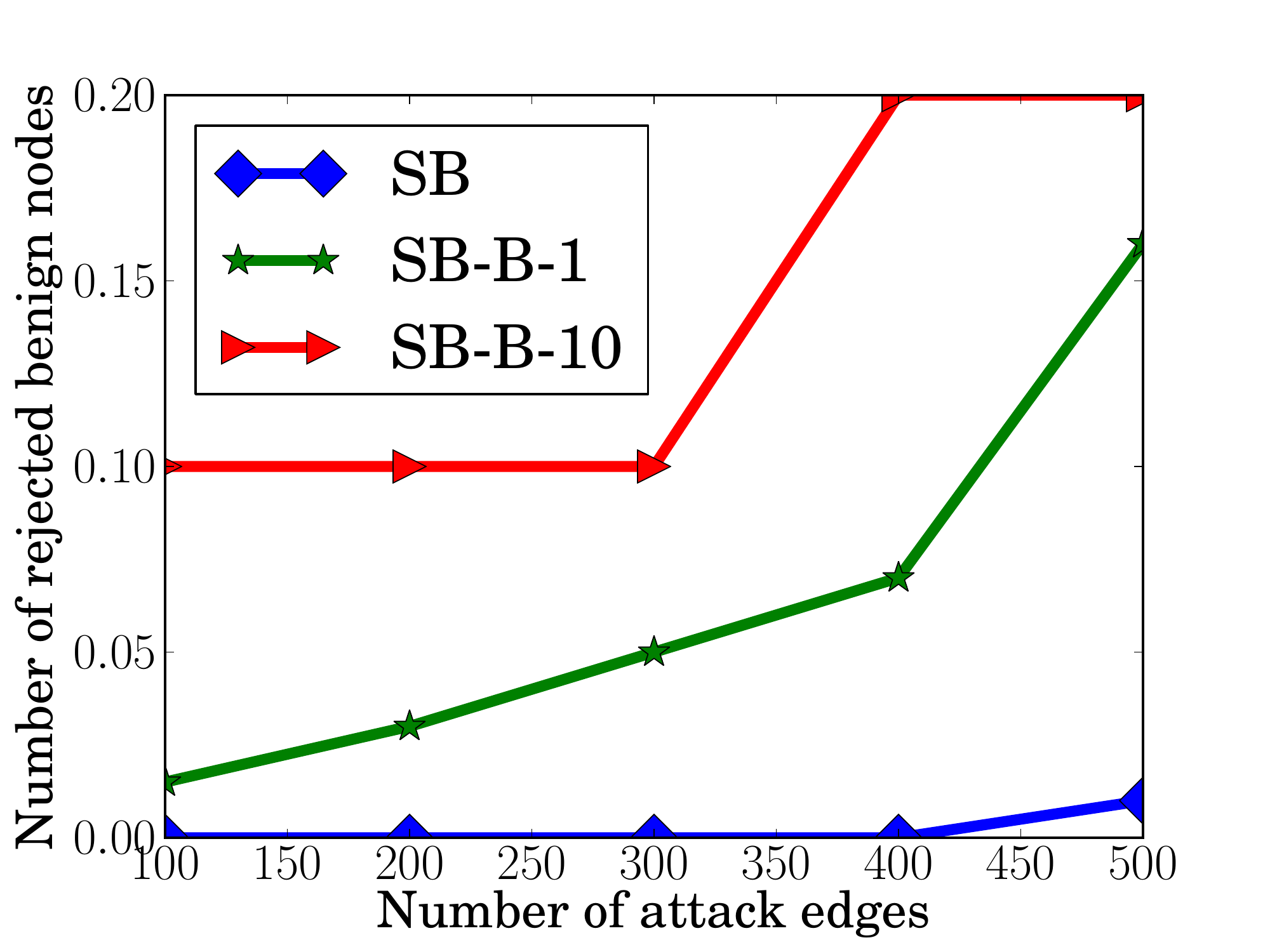}\label{explore_label_sampling_benign}}
\caption{Boosting with only labeled benign nodes. The numbers in the legend are the number of boosting trials. With only labeled benign nodes, SybilBelief (SB) with boosting can achieve performances comparable 
to the case where both benign and Sybil labels are observed. Furthermore, the number of boosting 
trials balances between the accepted Sybil nodes and rejected benign nodes.}
\label{explore_label_sampling}
\vspace{-4mm}
\end{figure}

\myparatight{Boosting with only labeled benign or Sybil nodes}
If we only observe labeled benign or Sybil nodes, we can boost our algorithm by sampling some nodes 
and treating them as labeled Sybil or benign nodes. Since only observing labeled benign nodes is 
symmetric to only observing labeled Sybil nodes in the algorithmic perspective, 
we take the former case as an example to illustrate our boosting strategy.

In these experiments, we assume 100 benign labels are observed but no Sybil labels are obtained. 
So we sample 10 nodes uniformly at random, and treat them as the Sybil labels. 
Other factors are fixed to be the natural settings. With a given benign region, a 
Sybil region, attack edges between them and the benign labels, we repeat this 
Sybil labels sampling process for $k$ times, and ensemble their results in the way 
described in our algorithm section. We use SB-B-k to denote the case where we use 
boosting strategy and the number of boosting trials is $k$. Figure~\ref{explore_label_sampling} 
compares SB-B-1, SB-B-10 and SB. In the setting of SybilBelief, we assume 100 benign labels and 10 Sybil labels. We conclude that with only partial labels, our boosting 
strategy can still achieve performance comparable to the scenario where both benign and 
Sybil labels are observed. Furthermore, the number of boosting trials balances between the 
accepted Sybil nodes and rejected benign nodes. More specifically, boosting with more 
trials accepts fewer Sybil nodes but rejects more benign nodes.  

\myparatight{Summary} We have the following observations:
\begin{itemize}
\item SybilBelief accepts more Sybil nodes in the PA-generated networks than in the 
ER-generated networks. This implies that attackers should design their Sybil 
regions to approach scale-free networks. 
\item SybilBelief is robust to label sites.
\item There exists a phase transition point $w_0$ (e.g., $w_0\approx 0.65$ in our experiments) for the parameter $w$. 
SybilBelief performance is robust for $w>w_0$. 
\item SybilBelief only requires one label per community.
\item SybilBelief can tolerate 49\% of labels to be incorrect. Moreover, SybilBelief can detect incorrect labels with 100\% accuracy.
\item SybilBelief is robust to community structures in the benign region.
\item With only benign or Sybil labels, our boosting strategy can still achieve performances comparable 
to the case where both benign and Sybil labels are observed. Furthermore, the number of boosting 
trials can be used to balance between accepted Sybil nodes and rejected benign nodes. 

\end{itemize}

\section{Comparing SybilBelief with Previous Approaches}
\label{eva_syn}

We compare our approach SybilBelief and its variants with previous Sybil classification and Sybil ranking mechanisms. The benign regions are real social networks while the Sybil regions are either generated by network generators such as Preferential Attachement~\cite{Barabasi99} or duplicates of the benign regions\footnote{{We acknowledge that one limitation of our work is that we didn't evaluate these approaches using real Sybil users. This is because it is hard for us to obtain a social network 
which represents trust relationships between users and which includes ground truth information about benign and Sybil users.}}. We find that SybilBelief and its variants perform orders of magnitude better than previous Sybil classification systems and significantly better than previous Sybil ranking systems. Furthermore, in contrast to previous approaches, SybilBelief is robust to label noise.

\begin{table}[!t]\renewcommand{\arraystretch}{1.3}
\centering
\caption{Dataset statistics. }
\centering
\begin{tabular}{|c|c|c|c|} \hline 
{\small Metric} & {\small Facebook} & {\small Slashdot} &{\small Email}\\ \hline
{\small Nodes} & {\small 43,953} & {\small 82,168} & {\small 224,832}\\ \hline
{\small Edges} & {\small 182,384} & {\small 504, 230} & {\small 339,925}\\ \hline
{\small Ave. degree} &  {\small 8.29} & {\small 12.27} & {\small 3.02}\\ \hline
\end{tabular}
\label{dataset}
\end{table}

\begin{figure}[t]
\vspace{-4mm}
\centering
\subfloat[Accepted Sybil nodes]{\includegraphics[width=0.25\textwidth, height=1.5in]{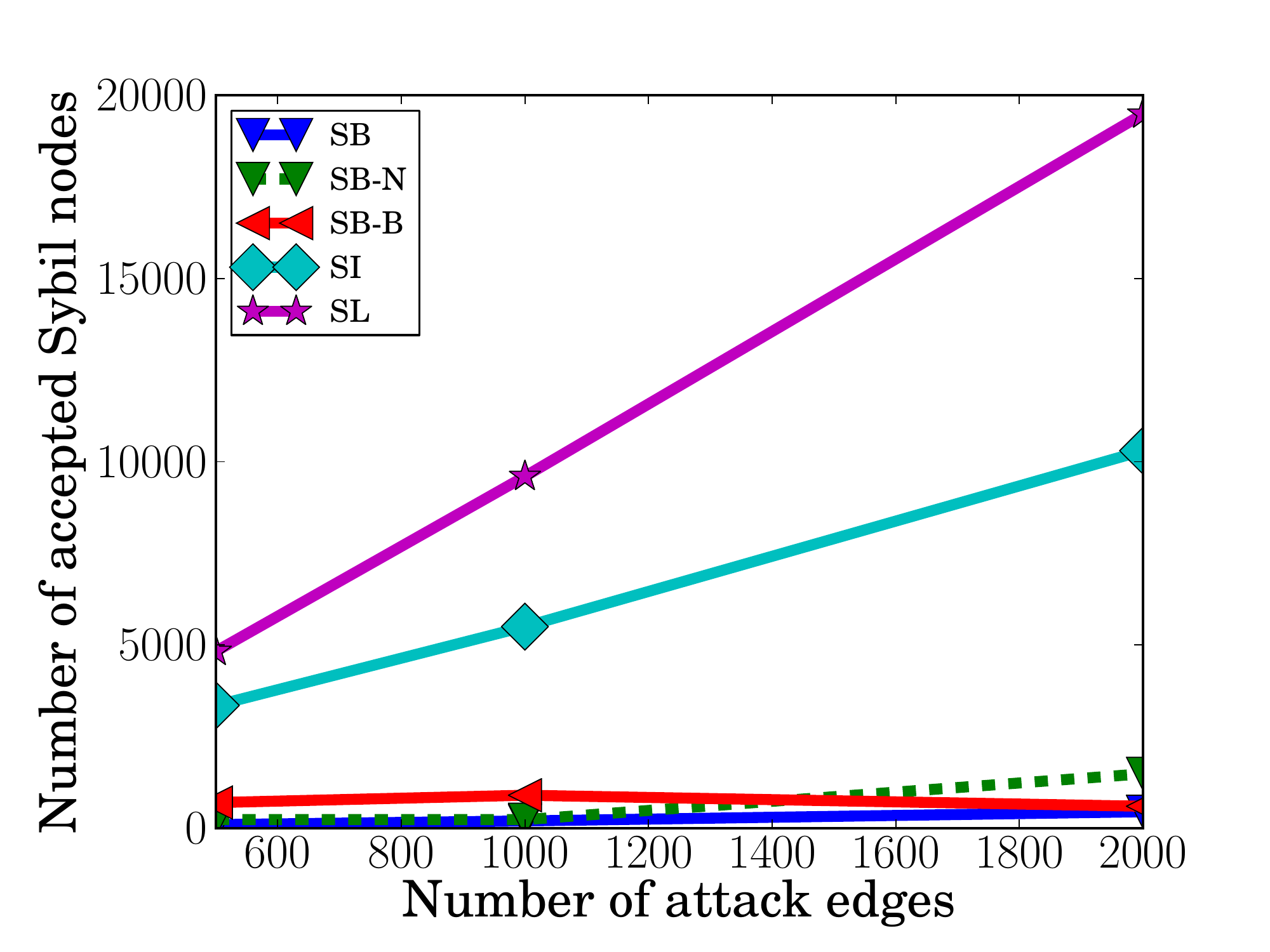}\label{sb-facebook-sybil}}
\subfloat[Rejected benign nodes]{\includegraphics[width=0.25\textwidth, height=1.5in]{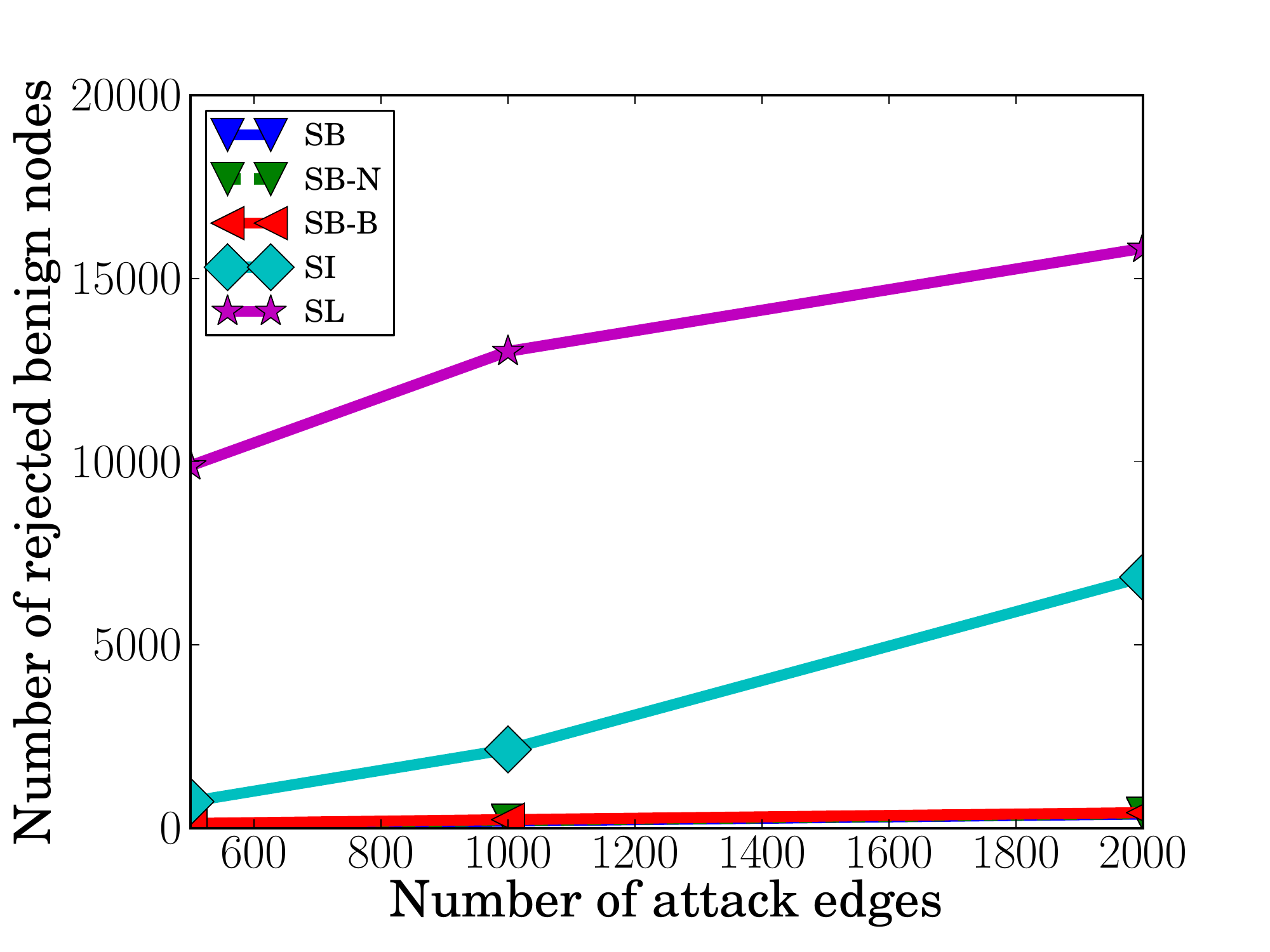}\label{sb-facebook-benign}}
\caption{The number of accepted Sybil nodes and rejected benign nodes as a
function of the number of attack edges. The benign region is the Facebook
network, and the Sybil regions are synthesized by PA model. We observe that SB,
SB-N, and SB-B all work an order of magnitude better than previous
classification systems. Furthermore, we find that incorporating both benign and Sybil labels
increases the performance of our algorithm.}
\label{sb-sybil}
\vspace{-4mm}
\end{figure}

\subsection{Experimental Setups}
\myparatight{Dataset description} We use three social networks representing different application scenarios. These three datasets are denoted as Facebook, Slashdot, and Email.  

Facebook is an interaction graph from the New Orleans regional network~\cite{VISWANATH09}. In this graph, nodes 
are Facebook users and a link is added between two users if they comment on each other's wall posts at least once. 
The Email network was generated using email data from a large European research institution~\cite{Leskovec05}. Nodes 
are email addresses and there exists a link between two nodes if they communicate with each other at least once. 
Slashdot is a technology-related news website, which allows users to tag each other as friends or foes. 
The Slashdot network thus contains friend/foe links between the users. We choose the largest connected component from each of them in our experiments. Table~\ref{dataset} summarizes the basic dataset statistics.

We note that some previous work removes nodes with degrees smaller than a threshold from the social networks. For instance,  SybilLimit~\cite{Yu08} removes nodes with degree smaller than 5 and SybilInfer~\cite{Danezis09} removes nodes with degree smaller than 3. Mohaisen et al.~\cite{mohaisen:imc10} found that such preprocessing will prune a large portion of nodes.  Indeed,  social networks often have a long-tail degree distribution (e.g., power-law degree distribution~\cite{Clauset09} and lognormal degree distribution~\cite{Gong12-imc}), in which most nodes have very small degrees. Thus,  a large portion of nodes are pruned by such preprocessing.   

Therefore, such preprocessing  could result in high false negative rates or high false positive rates depending on how we treat the pruned nodes. If we treat all the pruned nodes whose degrees are smaller than a threshold  as benign nodes, then an attacker can create many Sybil nodes with degree smaller than the threshold, resulting in high false negative rates, otherwise a large fraction of benign nodes will be treated as Sybil nodes, resulting in high false positive rates. So we do not perform such preprocessing to the three social networks.

\myparatight{Metrics} Following previous work~\cite{Danezis09,Yu08,Tran09,Tran11}, the metrics we adopt for evaluating Sybil classification mechanisms  are the number of \emph{accepted Sybil nodes} and the number of \emph{rejected benign nodes}, which are the maximum number of accepted Sybils and the maximum number of rejected benign nodes for a given number of attack edges. To evaluate the Sybil ranking mechanisms, similar to previous work~\cite{Viswanath10,Cao12}, we adopt AUC, Area Under the Receiver Operating Characteristic (ROC) Curve. Given a ranking of a set of benign and Sybil nodes, AUC is the probability that a randomly selected benign node ranks before a randomly selected Sybil node.  We compute the AUC in the manner described in~\cite{Hand01}.

\myparatight{Parameter settings} We set $\theta_v=0.5$ for any unlabeled node, which means we don't distinguish a node between benign and Sybil if no prior information is available. For those labeled nodes, their states are fixed and thus their corresponding parameters $\theta_v$ don't influence our model. So we also set them to be 0.5 by simplicity. Furthermore, as we find that SybilBelief is robust to the parameter $w_{uv}$ when it is bigger than the transition point, and thus we set $w_{uv}=w=0.90$ for any edge $(u,v)$ in all experiments. In the experiments of boosting our SybilBelief with only labeled benign nodes,  the number of boosting trials is set to be 10.

\subsection{Compared Approaches} 
We compare SybilBelief with two classical Sybil classification mechanisms, i.e., SybilLimit~\cite{Yu08} and SybilInfer~\cite{Danezis09}, and two recent Sybil ranking mechanisms, i.e., SybilRank~\cite{Cao12} and  Criminal account Inference Algorithm~\cite{Yang12-spam}. Table~\ref{alg} summarizes the notations of these algorithms. Note that Viswanath et al.~\cite{Viswanath10} showed that SybilLimit and SybilInfer are essentially also ranking systems. However,  Cao et al.~\cite{Cao12} showed that SybilRank outperforms them in terms of rankings. Thus, we will compare SybilBelief with SybilLimit and SybilInfer via treating them as Sybil classification mechanisms. 

\myparatight{SybilLimit (SL)} SL requires each node to run a few random routes starting from themselves. 
For each pair of nodes $u$ and $v$, if $v$'s random routes have enough intersections with $u$'s and these intersections satisfy a 
balance constraint, then $u$ accepts $v$ as a benign node. In our experiments, we random sample enough such pairs of 
nodes to estimate the number of accepted Sybil nodes and the number of rejected benign nodes.  

\myparatight{SybilInfer (SI)} SI relies on a special random walk, i.e., the stationary distribution of this random 
walk is uniform. Given a set of random walk traces, SI infers the posterior probability of any node being benign. 
Note that SI can only incorporate one labeled benign node.

\myparatight{SybilRank (SR)} SR performs random walks starting from a set of benign users. Specifically, with $h$ labeled benign nodes, SR designs a 
special initial probability distribution over the nodes, i.e., probability $1/h$ for each of the labeled benign nodes and probability 0 for all other nodes, and SR iterates the random walk from this initial distribution for $\text{log}(n)$ iterations, where $n$ is the 
number of nodes in the system. It is well known that this random walk is biased to high-degree nodes. Thus, SybilRank normalizes the final probabilities of nodes by their degrees and uses the normalized probabilities to rank nodes. Note that SR can only incorporate benign labels.

\begin{table}[!t]\renewcommand{\arraystretch}{1.1}
\centering
\caption{Notations of algorithms.}
\centering
\begin{tabular}{|c|c|} \hline 
{\small Notation} & {\small Description}\\ \hline
{\small SL} & {\small SybilLimit~\cite{Yu08}}\\ \hline
{\small SI} & {\small SybilInfer~\cite{Danezis09}}\\ \hline
{\small SR} & {\small SybilRank~\cite{Cao12}}\\ \hline
{\small SR-N} & {\small SybilRank~\cite{Cao12} with label noise}\\ \hline
{\small CIA} & {\small Criminal account Inference Algorithm~\cite{Yang12-spam}}\\ \hline
{\small CIA-N} & {\small CIA with label noise}\\ \hline
{\small SB} & {\small SybilBelief}\\ \hline
{\small SB-N} & {\small SybilBelief with label noise}\\ \hline
{\small SB-B} & {\small SybilBelief with only labeled benign nodes}\\ \hline
{\multirow{2}{*}{\small Random}} & {\small Randomly assign a reputation score}\\
&{\small between 0 and 1 to each node}\\ \hline
\end{tabular}
\label{alg}
\end{table}

\myparatight{SybilRank-Noise (SR-N)} We use this abbreviation to denote the case where the labels given to SybilRank are noisy, i.e., some of the labeled 
benign nodes are actually Sybils.

\myparatight{Criminal account Inference Algorithm (CIA)} CIA is also based on a random walk with a special initial probability distribution, but it differs from SR in two major aspects. 
First, CIA starts the random walk from labeled Sybil nodes. Second, in each iteration, CIA restarts the random walk from the special 
initial probability distribution with probability $1-\alpha$. Since the random walk is started from Sybil nodes, 1 -  $p_v$ is treated as the reputation score of node $v$, where $p_v$ is the stationary probability of $v$. Then those reputation scores are used to rank nodes. As was proposed in the original paper~\cite{Yang12-spam}, we set the restart 
parameter $\alpha$ to  0.85. Note that CIA can only incorporate labeled Sybil nodes.

\myparatight{Criminal account Inference Algorithm-Noise (CIA-N)} Analogous to SR-N, we use this abbreviation to denote the case where the input labels are partially wrong.

We abbreviate variants of our method by {\em SybilBelief (SB), SybilBelief-Noise (SB-N) and SybilBelief-Boosting (SB-B)}. SB incorporates both benign and Sybil labels; SB-N indicates the scenario where some of the labeled benign and Sybil nodes are noise; SB-B means only benign  labels are observed, and we  sample some nodes uniformly at random from the entire network and treat them as Sybil labels.

\begin{figure*}[!t]
\vspace{-4mm}
\centering
\subfloat[Facebook]{\includegraphics[width=0.33\textwidth, height=2in]{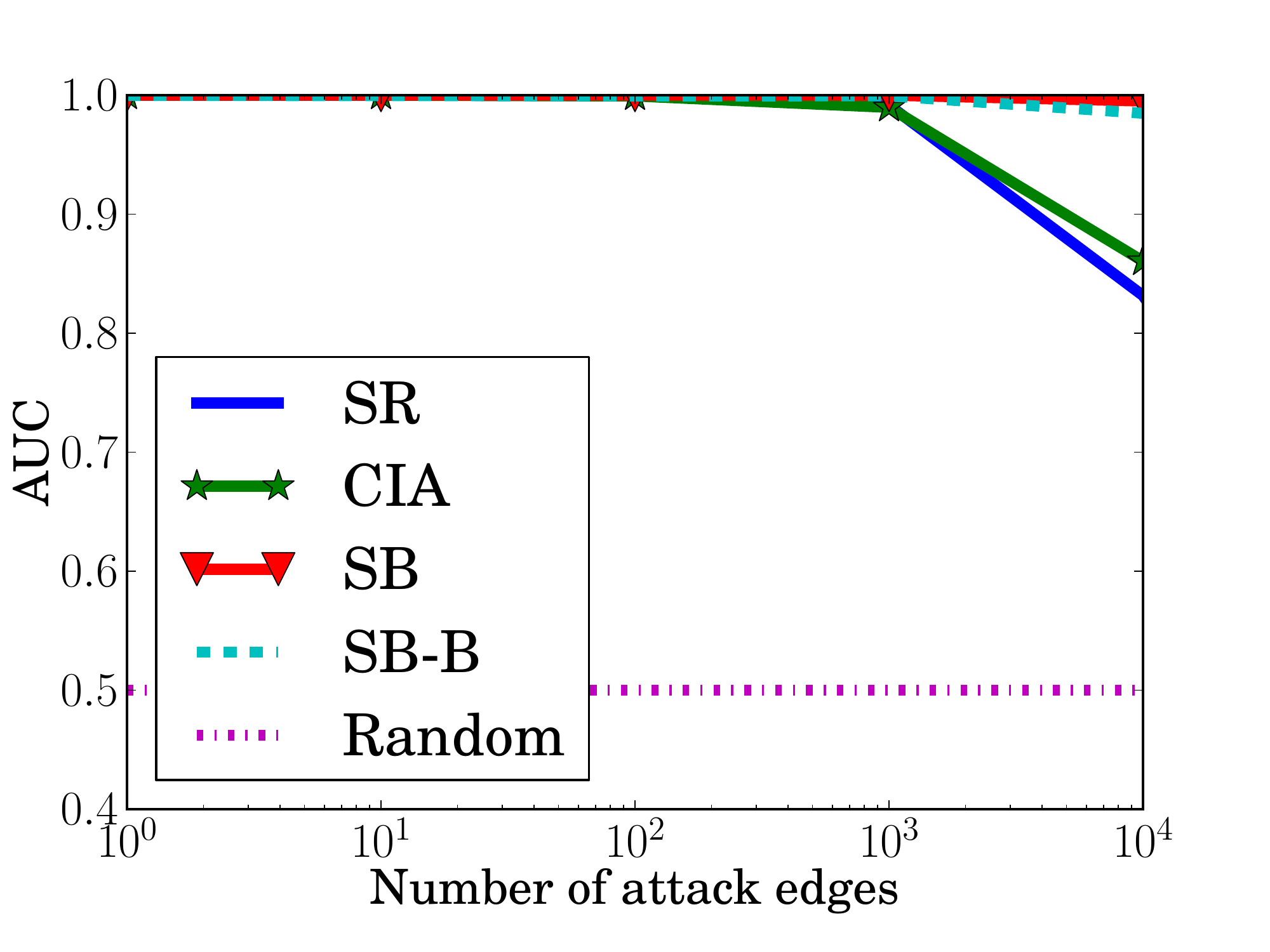}\label{sybilrank-facebook}}
\subfloat[Slashdot]{\includegraphics[width=0.33\textwidth, height=2in]{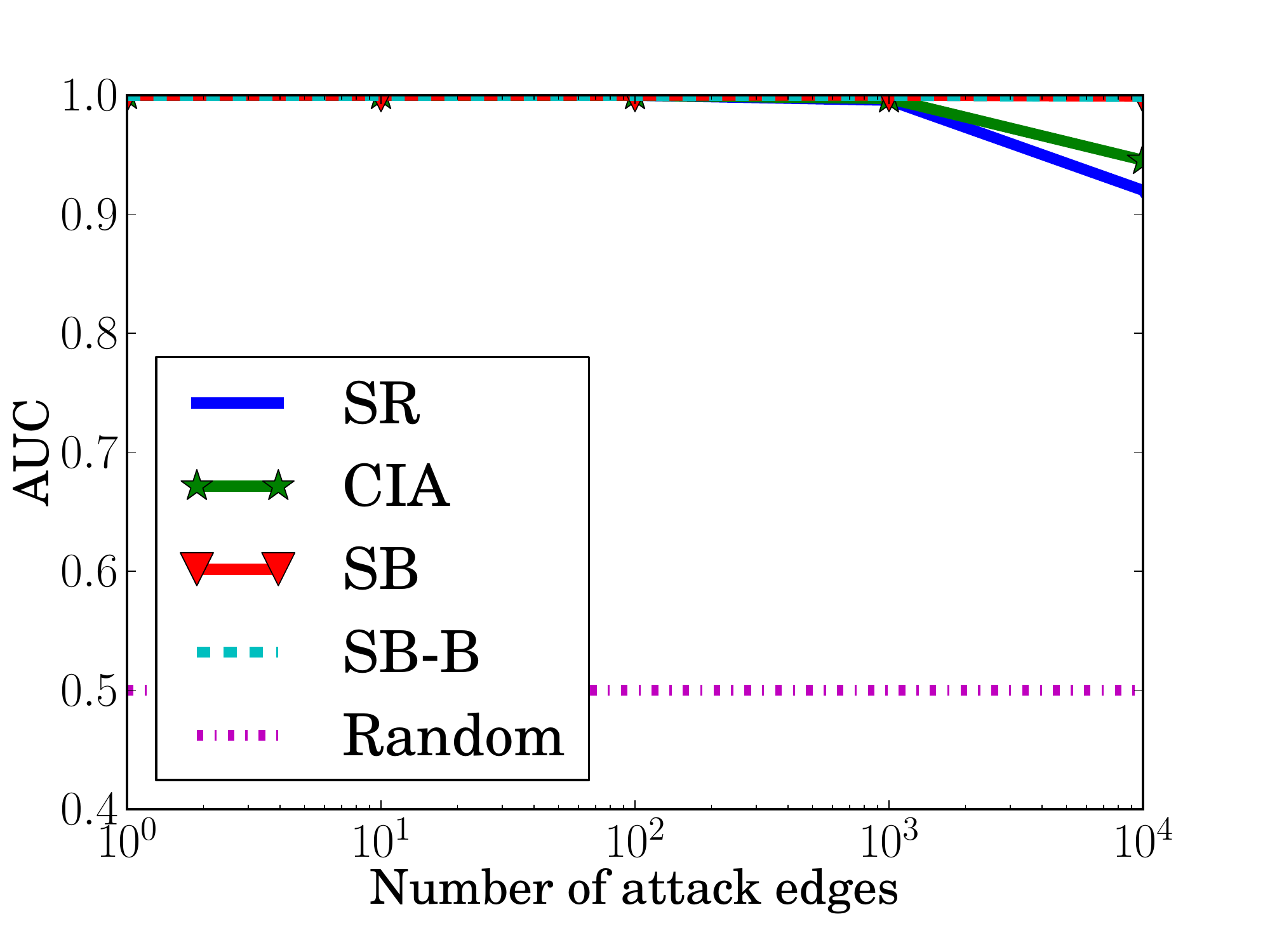}\label{sybilrank-slashdot}}
\subfloat[Email]{\includegraphics[width=0.33\textwidth, height=2in]{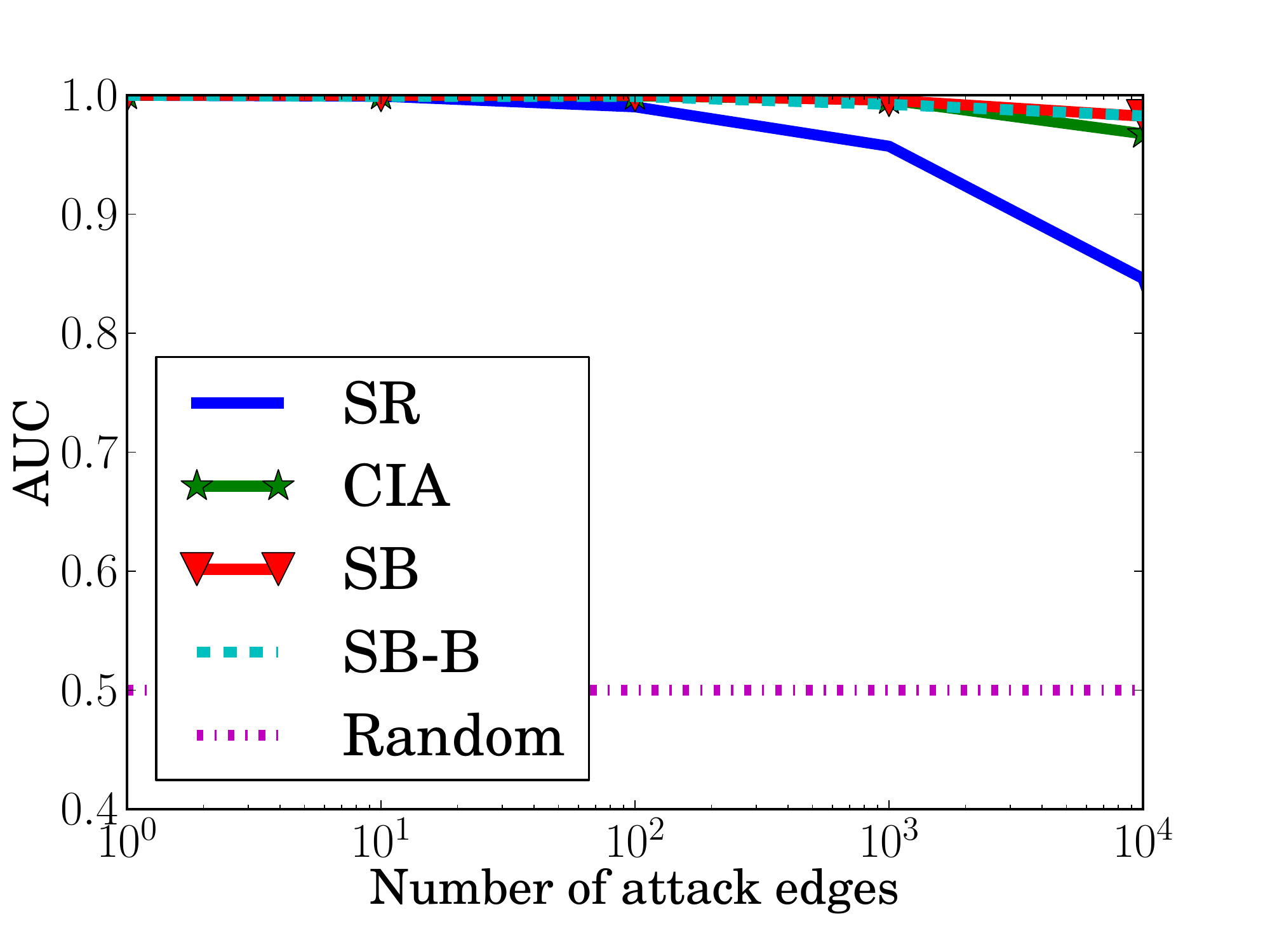}\label{sybilrank-email}}
\vspace{-2mm}

\subfloat[Facebook with label noise]{\includegraphics[width=0.33\textwidth, height=2in]{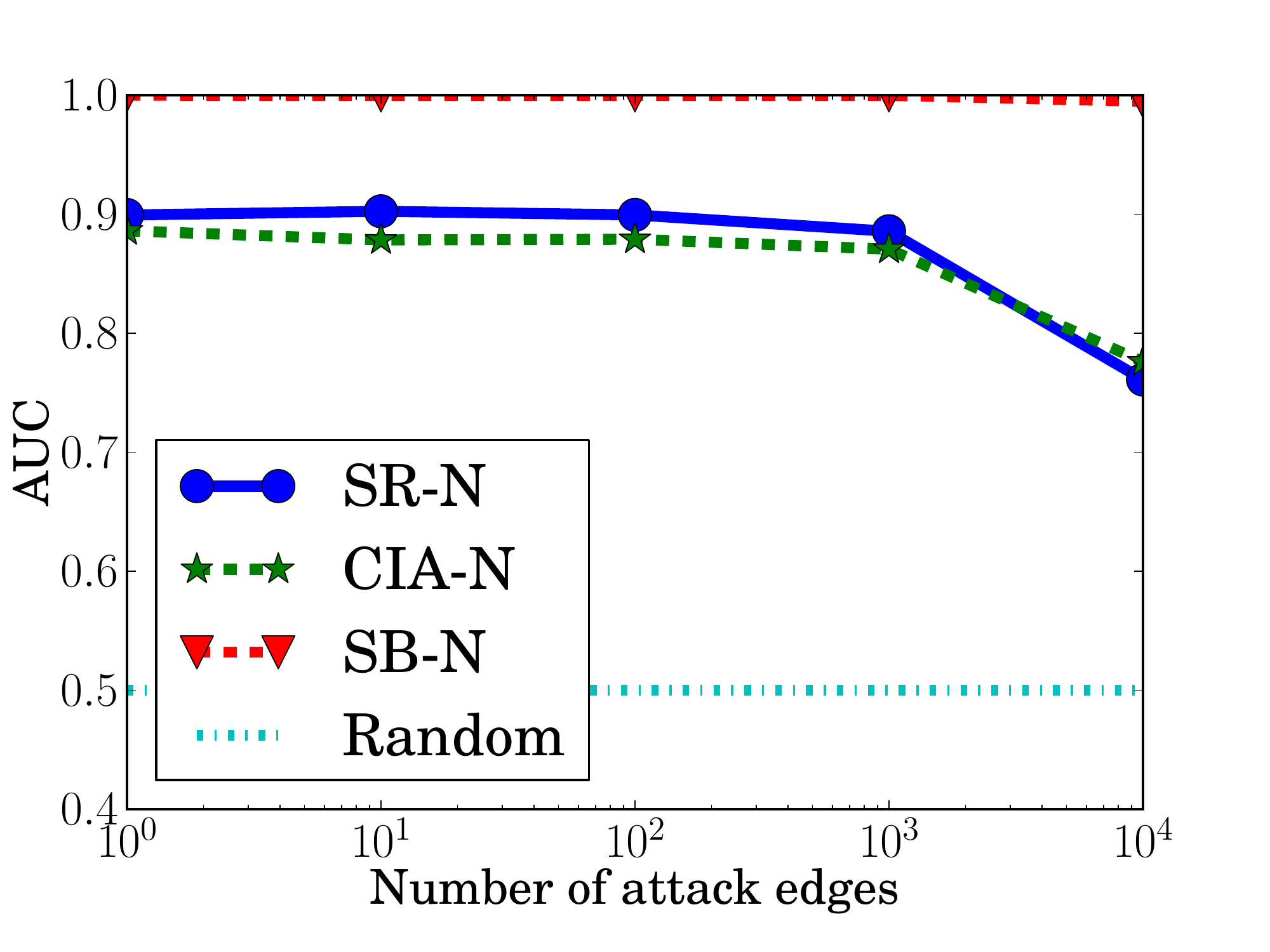}\label{sybilrank-facebook}}
\subfloat[Slashdot with label noise]{\includegraphics[width=0.33\textwidth, height=2in]{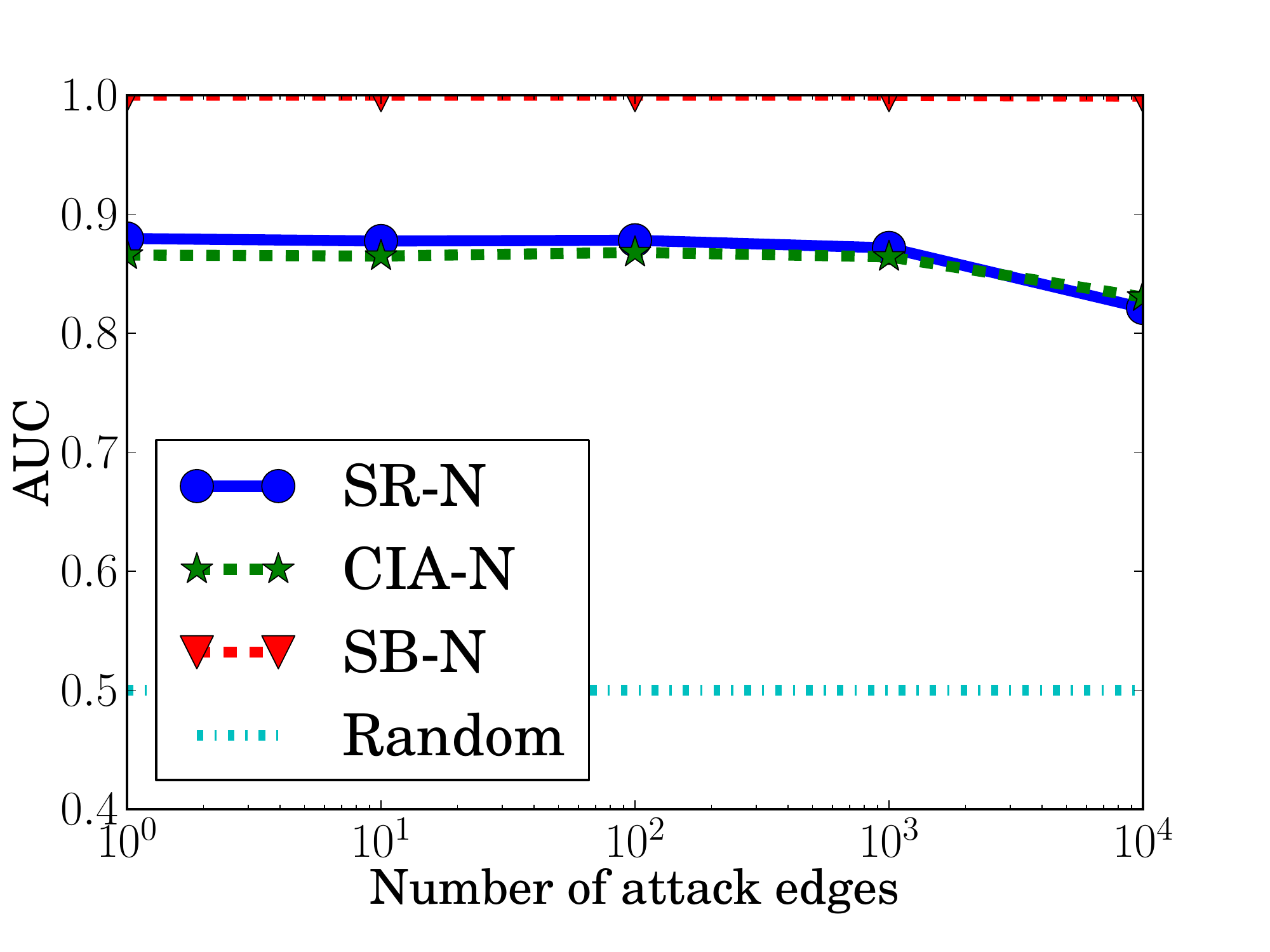}\label{sybilrank-slashdot}}
\subfloat[Email with label noise]{\includegraphics[width=0.33\textwidth, height=2in]{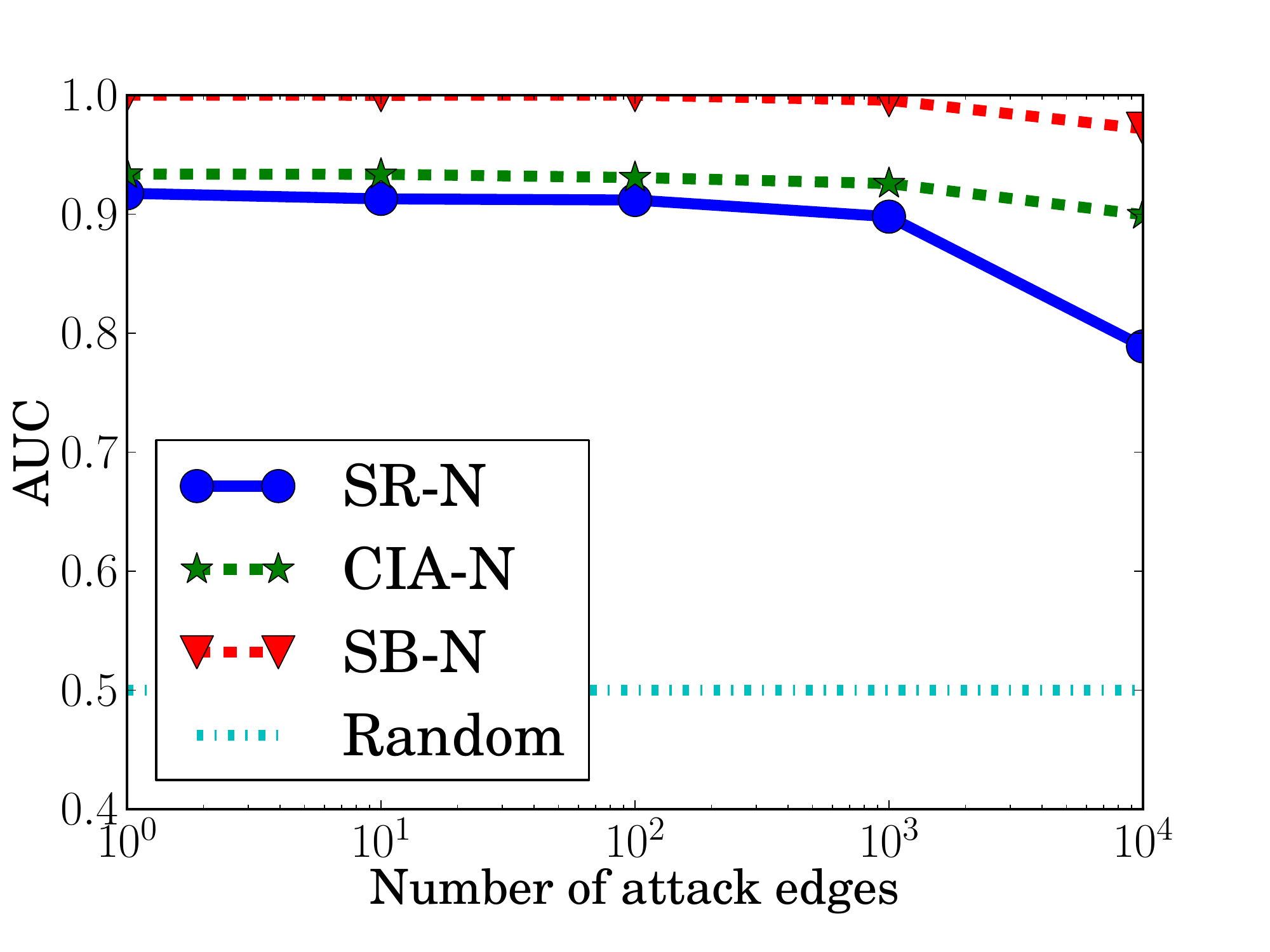}\label{sybilrank-email}}
\caption{AUC as a function of the number of attack edges on different social
networks. For each social network, we treat it as both the benign and Sybil
regions, and add attack edges between them uniformly at random. The AUCs are
averaged over 10 trials for each number of attack edges. We observe that both SB
and SB-B outperform previous approaches. Furthermore, in contrast to previous
approaches, SB is robust to label noise.}
\label{auc}
\vspace{-4mm}
\end{figure*}

\subsection{Comparing with Sybil Classification Mechanisms}  
In order to calculate the number of \emph{accepted Sybil nodes} and the number of \emph{rejected benign nodes}, we need to evaluate these algorithms on Sybil regions with various sizes. Thus, we use a network generator to synthesize the Sybil region and add attack edges between it and the benign region uniformly at random. In our experiments, we adopt scale-free network generator Preferential Attachment (PA)~\cite{Barabasi99}. However, we still use real social networks as the benign regions. We assume 100 labeled benign nodes, which are sampled from the benign region uniformly at random. Since SybilLimit and SybilInfer don't incorporate labeled Sybil nodes, we assume only one labeled Sybil node for our approaches, which is sampled from the Sybil region uniformly at random.  In the experiments with label noise, we assume 10 out of the labeled benign nodes are incorrect. In the boosting experiments, we randomly sample 10 nodes from the entire network and  treat them as Sybil labels.
   
Figure~\ref{sb-sybil} shows the comparison results on the Facebook dataset. We only show results on the Facebook network 
because SybilLimit and SybilInfer are not scalable to other social networks we consider. We have several observations. First, SybilBelief performs orders of magnitude better than SybilLimit and SybilInfer in  terms of both the number of accepted Sybil nodes and the number of rejected benign nodes. Second, unsurprisingly, SB-N and SB-B don't work as well as SB, but the margins are not significant.  Since SybilLimit and SybilInfer can only incorporate one labeled benign node, the gains of SB and SB-B over them come from (a) incorporating more labels and (b) the use of the Markov Random Fields and the Belief Propagation algorithm.

\subsection{Comparing with Sybil Ranking Mechanisms} Following previous work~\cite{Cao12,Viswanath10,Yu08,Danezis09,Alvisi13}, we synthesize networks as follows:
we use each real-world social networks as both the benign region and the Sybil region, and then we add attack edges between them uniformly at random. 

We assume that 100 labeled benign nodes and 100 labeled Sybil nodes are given. For the experiments with label noise, we assume 10 out of the 100 benign and Sybil labels are wrong.  For   SB-B, we randomly sample 100 nodes from the entire network including both benign and Sybil regions and treat them as Sybil labels.

Figure~\ref{auc} shows the AUC of the rankings obtained by various approaches in the three different social networks. From these figures, 
we can draw several conclusions. 

First, SB and SB-B consistently outperform SR and CIA across different social networks. Furthermore, the improvements are more significant as the number of attack edges becomes larger. This is because our proposal can incorporate both benign and Sybil labels. 

Second, with label noise, performances of both SR and CIA degrade dramatically. However, SB's performance is  almost unchanged. SB is robust to noise because SB incorporates labels probabilistically. Specifically, a node receives beliefs from all of its neighbors. Since the majority of labels are correct, the beliefs from the wrongly labeled nodes are dominated by those from the correctly labeled nodes. 

Third, CIA consistently performs better than SR, which can be explained by the fact that CIA restarts the random walk from the special initial probability distribution with some probability in each iteration. 

\subsection{Summary} We have performed extensive evaluations to compare our approach SybilBelief and its variants with previous approaches on graphs with synthetic Sybil nodes. From the comparison results, we conclude that SybilBelief and its variants perform orders of magnitude better than previous Sybil classification systems and significantly better than previous Sybil ranking systems. Furthermore, in contrast to previous approaches, SybilBelief is robust to label noise.

\section{Related Work}

\subsection{Structure-based Sybil Defenses}
Most existing structure-based Sybil defenses  are based on either random walks or community detections. We refer readers to a recent  survey~\cite{Alvisi13} for more details.
  
\myparatight{Random walk based Sybil classification}
SybilGuard~\cite{Yu06} and SybilLimit~\cite{Yu08} were the first 
schemes to propose Sybil detection using social network structure. SybilLimit 
relies on the insight that social networks are relatively well 
connected, and thus short random walks starting from benign 
users can quickly reach all other benign users~\footnote{More
precisely, SybilGuard and SybilLimit use a variant of random walks
 called random routes. Please see~\cite{Yu06,Yu08} for more detail.}. 
Thus if two benign users perform $\sqrt{m}$ random walks (where 
$m$ is the number of edges between benign users), then they will 
have an intersection with high probability, using the birthday 
paradox. The intersection of random walks is used as a feature 
by the benign users to validate each other. On the other hand, 
short random walks from Sybil users do not reach all benign 
users (due to limited number of attack edges), and thus do not 
intersect with the random walks from benign users. 

SybilInfer~\cite{Danezis09} aims to directly detect a bottleneck 
cut between benign and Sybil users. SybilInfer relies on random 
walks and uses a combination of Bayesian inference and Monte-Carlo 
sampling techniques to estimate the set of benign and Sybil users.  
%
%
Similar to SybilGuard, SybilLimit and SybilInfer, Gatekeeper~\cite{Tran11} 
and SybilDefender~\cite{sybildefender} also leverage random walks. These mechanisms 
make additional assumptions about the structure of benign and Sybil nodes, 
and even the size of the Sybil population~\cite{sybildefender}.  Moreover, they also require the benign regions to be fast mixing, which was shown to be unsatisfied by  Mohaisen et al.~\cite{mohaisen:imc10}.


In contrast to the above approaches, SybilBelief does not use random walks, 
and relies instead on the Markov Random Fields and Loopy Belief Propagation. SybilBelief 
is able to incorporate information about known benign and known Sybil nodes. 
Our experimental results show that SybilBelief performs an order of magnitude 
better than SybilLimit and SybilInfer. Moreover, SybilBelief is scalable to large scale social networks, unlike above mechanisms.



\myparatight{Random walk based Sybil ranking}
SybilRank~\cite{Cao12}  performs random walks starting from a set of benign users. Specifically, with $h$ labeled benign nodes, SybilRank designs a 
special initial probability distribution over the nodes, i.e., probability $1/h$ for each of the labeled benign nodes and probability 0 for all other nodes, and iterates the random walk from this initial distribution for $\text{log}(n)$ iterations, where $n$ is the total 
number of nodes in the network. It is well known that this random walk is biased to high-degree nodes. Thus, SybilRank normalizes the final probabilities of nodes by their degrees and uses the normalized probabilities to rank nodes. SybilRank scales to very large social networks
and has shown good performance on the Tuenti network. However, 
SybilRank has two major limitations: (a) it does not tolerate  label noise, and (b) it does not incorporate Sybil labels. 

CIA is also based on a random walk with a special initial probability distribution, but it differs from SybilRank in two major aspects. 
First, CIA starts the random walk from labeled malicious nodes. Second, in each iteration, CIA restarts the random walk from the special 
initial probability distribution with some probability. Since the random walk is started from malicious nodes, 1 -  $p_v$ is treated as the reputation score of node $v$, where $p_v$ is the stationary probability of $v$. Then those reputation scores are used to rank nodes.  CIA  scales well  to large OSNs. However, CIA is unable to incorporate known benign labels and thus is fundamentally inapplicable in settings where  a set of Sybil labels are unavailable. We found that CIA is also not resilient 
to label noise. 


Alvisi et al.~\cite{Alvisi13} studied Personalized PageRank (PPP) as a Sybil ranking mechanism. Unlike SybilRank, PPP incorporates only one benign label by initiating a random walk from the labeled benign node and returning back to it in each step  with some probability. Similar to SybilRank, the normalized stationary probability distribution of PPP is used to rank all the users.  
SybilRank is better than PPP because  PPP only incorporates one benign label. 

These random walk based Sybil ranking approaches have complexity of $O(n\text{log}n)$~\cite{Cao12}, where $n$ is the number of nodes in the social network. Our SybilBelief has a complexity of $O(nd)$, where $d$ is the number of iterations. In practice, $\text{log}n$ and $d$ are very similar.  In our experiments, we compared SybilBelief with SybilRank and CIA. We found that their computation times are similar.

\myparatight{Community detection based Sybil classification}\\
Viswanath et al.~\cite{Viswanath10} showed that the Sybil detection problem 
can be cast as a community detection problem. In their experimental evaluation, 
the authors found that using a simple local community detection algorithm proposed by 
Mislove et al.~\cite{mislove10} had equivalent results to using the state-of-art 
Sybil detection approaches. However, their scheme has computational complexity as high as
$O(n^2)$, and thus does not scale to multi-million node social networks. Their approach is 
also unable to simultaneously incorporate both known benign and Sybil nodes. Moreover, Alvisi et al.~\cite{Alvisi13} showed that their local community detection algorithm is not robust to advanced attacks by constructing such an attack.

Cai and Jermaine~\cite{cai:ndss12} proposed to detect Sybils using a latent community 
detection algorithm. With a hierarchical generative model for the observed social network, 
detecting Sybils is mapped to an Bayesian inference problem. Cai and Jermaine adopted Gibbs 
sampling, an instance of Markov chain Monte Carlo (MCMC) method, to perform the inference. 
However, it is well known in {the} machine learning community that {the} 
MCMC method is not scalable.

\subsection{Trust Propagation}
Another line of research, which is closely related to Sybil detection, is the trust propagation problem. Several approaches  have been proposed to propagate trust scores or reputation scores in file-sharing networks or auction platforms 
(e.g.,~\cite{Kamvar:2003:EAR:775152.775242,Richardson03trustmanagement,Guha:2004:PTD:988672.988727}). 
Similar to SybilRank~\cite{Cao12} and CIA~\cite{Yang12-spam}, these approaches are variants of the PageRank algorithm~\cite{Brin98}. 
In principle, these approaches can be applied to detect Sybils. Specifically, nodes with low trust scores could be classified as Sybils~\cite{Cao12}.  

\subsection{Markov Random Fields and Belief Propagation}

The Markov Random Fields (MRF) has many applications in  electrical engineering and computer science  such as computer vision~\cite{Cross83} and natural language processing~\cite{Manning99}.  However, the application of MRFs to the security and privacy area is rather limited.

Computing posterior  distributions in probabilistic graphical models is a central problem in 
probabilistic reasoning. It was shown that this problem is NP-hard on general graphs~\cite{Cooper:1990:CCP:77754.77762}.
Pearl proposed belief propagation algorithm for exact inference on trees, 
and he also noted that, on graphs with loops, this algorithm leads to oscillations that prohibit 
any convergence guarantees~\cite{Pearl88}. Nevertheless, belief propagation on loopy graphs 
has often been used in practical applications and has demonstrated satisfying approximate performance~\cite{Murphy99}.

\section{Conclusion}
In this paper, we propose SybilBelief, a semi-supervised learning framework, to detect Sybil nodes in distributed systems. SybilBelief takes social networks among the nodes in the system, a small set of known benign nodes, and, optionally, a small set of known Sybil nodes as input, and then SybilBelief propagates the label information from the known benign and/or Sybil nodes to the remaining ones in the system.

We extensively evaluate the influence of various factors including parameter settings in the SybilBelief, the number of labels, and label noises on the performance of SybilBelief. Moreover, we compare SybilBelief with state-of-the-art Sybil classification and ranking approaches on real-world social network topologies. Our results demonstrate that SybilBelief performs orders of magnitude better than previous Sybil classification mechanisms and significantly better than previous Sybil ranking mechanisms. Furthermore, SybilBelief is more resilient to noise in our prior knowledge about known benign nodes and known Sybils.

Interesting avenues for future work include evaluating SybilBelief and previous approaches with datasets containing real Sybils and applying our SybilBelief framework to other security 
and privacy problems such as graph based Botnet detection~\cite{botgrep}, reputation systems~\cite{Kamvar:2003:EAR:775152.775242}, and private information inference~\cite{Gong11}.

\end{document}